\documentclass[11pt]{article}
\usepackage[utf8]{inputenc}
\usepackage{graphicx}
\usepackage{amsmath,amssymb,bm}

\usepackage[body={6in,8.5in}]{geometry}

\newcommand {\unitv}[1]{\mathbf{\hat{#1}}}

\newcommand {\zhat}{\unitv{z}}
\newcommand {\yhat}{\unitv{y}}
\newcommand {\xhat}{\unitv{x}}

\newcommand{\beq}{\begin{eqnarray}}
\newcommand{\eeq}{\end{eqnarray}}
\newcommand{\bi}{\begin{itemize}}
\newcommand{\ei}{\end{itemize}}
\newcommand{\be}{\begin{enumerate}}
\newcommand{\ee}{\end{enumerate}}
\newcommand{\nin}{\noindent}

\newcommand{\defn}{\equiv}
\newcommand{\tit}{\textit}
\newcommand{\tbf}{\textbf}
\newcommand{\eps}{\mathrm{Ro}}
\newcommand{\Bu}{\mathrm{Bu}}
\newcommand{\Ro}{\mathrm{Ro}}
\newcommand{\scc}{s^\ast}
\newcommand{\U}{\mathcal{U}}
\newcommand{\grad}{\bm{\nabla}}
\newcommand{\curl}{\bm{\nabla\times}}
\newcommand{\divg}{\bm{\nabla\cdot}}

\title{Reduced models for wave-balanced flow interactions}
\author{André Palóczy}
\date{\today $\;$ (revised version)}

\begin{document}

\maketitle


\section{Introduction and goals}
\label{intro}

Energy pathways in the ocean encompass the full range of scales from planetary, O(10$^7$ m) down to microscales, O(10$^{-2}$ m). Energy is introduced in the ocean mostly at seasonal, planetary scales by the time-mean atmospheric forcing; at diurnal, planetary scales by the most energetic modes of the barotropic tide; and in the weather band (days to weeks) by the synoptic weather systems. This energy must either leave the ocean or be converted into internal energy at the Kolmogorov scale, O(1 cm) \cite{ferrari_wunsch2010}.

The routes mechanical energy takes from injection to dissipation in the ocean are currently an open problem. A group of candidates that has been gaining attention over the past decade includes processes involving the interaction of near-inertial waves (NIWs) with the mesoscale balanced flow, \textit{e.g.}, \cite{young_benjelloul1997,xie_vanneste2015,rocha_etal2018}. Two broad classes of such processes can be identified: The first is generation of NIWs by mesoscale and submesoscale instabilities, oftentimes called \tit{spontaneous} loss of balance. The second is the interaction of the balanced flow with existing NIWs by processes such as refraction, advection and dispersion, a mechanism that has been more studied in recent years and has been called \tit{stimulated} loss of balance, \tit{e.g.}, \cite{rocha_etal2018}.

The goals of this project are to: \tbf{1)} Study the energy exchanges between NIWs and the balanced flow using idealized simulations and \tbf{2)} To derive a new asymptotic model to help develop an understanding of such interactions.

\section{Non-asymptotic reduced models}

\subsection{The modified Thomas \& Yamada (2019) model}

We develop a new reduced model consisting of the barotropic mode and a high baroclinic mode, which we denote as T-W model hereafter (Section \ref{sec:TW} below). This model is a further reduction of the models implemented by \cite{thomas_yamada2019} and \cite{thomas_arun2020}. We begin by generally following \cite{thomas_yamada2019}'s derivation of the parent model to the T-W model. The starting point is the hydrostatic Boussinesq equations, with the assumption of constant buoyancy frequency $N$:

\beq
\bm{u}_t + \bm{u\cdot\nabla u} + w\bm{u}_z + \bm{f\times u} + \grad p = 0,\label{bouss1}\\
p_z = b,\\
b_t + \bm{u\cdot\nabla}b + wN^2 = 0,\\
\divg\bm{u} + w_z = 0,\label{bouss4}
\eeq

\nin where $(u, v, w)$ are the velocity components of $\bm{u}$ in the $(\bm{\hat{x}}, \bm{\hat{y}}, \bm{\hat{z}})$ directions, respectively, $p$ is pressure (normalized by a reference density $\rho_0$), $b \equiv -g\rho'/\rho_0$ is the buoyancy (where $\rho'$ is the perturbation pressure), $\bm{\nabla} \equiv\bm{\hat{x}}\partial_x + \bm{\hat{y}}\partial_y$ is the horizontal gradient operator, $\bm{f} \equiv \bm{\hat{z}}f$ (where $f$ is the inertial frequency) and $N$ is the buoyancy frequency. We expand all variables in the following form:

\beq
\bm{u}(\bm{x}, z, t) = \bm{u_0}(\bm{x}, t) + \sum_{n=1}^\infty\bm{u}(\bm{x},t)\phi'_n(z)\label{sl1}\\
w(\bm{x}, z, t) = \sum_{n=1}^\infty w(\bm{x},t)\phi_n(z)\label{sl2}\\
p(\bm{x}, z, t) = p_0(\bm{x}, t) + \sum_{n=1}^\infty\lambda_n^{-2}p_n(\bm{x},t)\phi'_n(z)\label{sl3}\\
b(\bm{x}, z, t) = -\sum_{n=1}^\infty p_n(\bm{x},t)N^2\phi_n(z)\label{sl4}
\eeq

\nin where $\phi_{n=0}(z) = 1$, $\phi_{n>0} = \sin(n\pi z)$ is the solution of the Sturm-Liouville problem with constant stratification $N(z) = 1$ and rigid lid boundary conditions, \tit{i.e.},

\beq
\phi''_n + \lambda_n^2N^2\phi_n = 0, \;\;\;\; \text{with} \;\;\;\; \phi_n(0) = \phi_n(1) = 0
\eeq

\nin where the eigenvalues are $\lambda_{n=0} = 0$ and $\lambda_{n>0} = n\pi$.

Restricting \ref{sl1}-\ref{sl4} to the barotropic mode (subscript $T$) and the $n$-th baroclinic mode (subscript $C$) gives

\beq
\bm{u}(\bm{x},t), p(\bm{x},t) = [\bm{u_T}(\bm{x},t), p_T(\bm{x},t)] + [\bm{u_C}(\bm{x},t), p_C(\bm{x},t)]\times\sqrt{2}\cos\bigg(\frac{n\pi z}{H}\bigg)\label{modes1}\\
w(\bm{x},t), b(\bm{x},t) = [w_C(\bm{x},t), b_C(\bm{x},t)]\times\sqrt{2}\sin\bigg(\frac{n\pi z}{H}\bigg),\label{modes2}
\eeq

\nin Substituting \ref{modes1}-\ref{modes2} into \ref{bouss1}-\ref{bouss4} and using the orthogonality property of the modes results in equations similar to the linear shallow water equations for the $n$-th baroclinic mode ($n \geq 1$):

\beq
\partial_t\bm{u_T} + \bm{f\times u_T} + \bm{\nabla}p_T + \Ro\big[\bm{u_T\cdot\nabla u_T} + \bm{u_C\cdot\nabla u_C} + \big(\bm{\nabla\cdot u_C}\big)\bm{u_C}\big] = 0,\label{TY1}\\
\bm{\nabla\cdot u_T} = 0,\\
\partial_t\bm{u_C} + \bm{f\times u_C} + \bm{\nabla}p_C + \Ro\big(\bm{u_T\cdot\nabla u_C} + \bm{u_C\cdot\nabla u_T}\big) = 0,\label{TY3}\\
\partial_tp_C + \bigg(\frac{NH}{n\pi}\bigg)^2\bm{\nabla\cdot u_C} + \Ro\big(\bm{u_T\cdot\nabla}p_C\big) = 0,\label{TY4}
\eeq

\nin where the Rossby number is (with characteristic velocity and horizontal length scales $U$ and $L$, respectively)

\beq
\Ro \equiv \frac{U}{fL}.
\eeq

\nin Taking curl of the T-mode's momentum equation \ref{TY1} to eliminate $p_T$,

\beq\label{Tvort}
\partial_t\zeta_T + \Ro\bm{\nabla\times}\big[\bm{u_T\cdot\nabla u_T} + \bm{u_C\cdot\nabla u_C} + (\bm{\nabla\cdot u_C})\bm{u_C}\big] = 0,
\eeq

\nin where $\zeta_T \equiv \partial_xv_T - \partial_yu_T$. Rescaling the baroclinic pressure as $p_C \to \Bu_np_C$ gives the final set of equations:

\beq
\partial_t\zeta_T + \Ro\bm{\nabla\times}\big[\bm{u_T}\cdot\bm{\nabla} \bm{u_T} + \bm{u_C}\cdot\bm{\nabla} \bm{u_C} + (\bm{\nabla\cdot u_C})\bm{u_C}\big] = 0,\label{ty11}\\
\partial_t\bm{u_C} + \bm{\hat{z}\times u_C} + \Bu_n\grad p_C + \Ro\big(\bm{u_T\cdot\nabla u_C} + \bm{u_C\cdot\nabla u_T}\big) = 0,\label{ty22}\\
\partial_tp_C + \divg \bm{u_C} + \Ro\big(\bm{u_T\cdot\nabla}p_C\big) = 0,\label{ty33}
\eeq

\nin where the modal Burger number is (with a characteristic vertical length scale $H$)

\beq
\Bu_n \equiv \bigg(\frac{NH}{\lambda_n fL}\bigg)^2
\eeq

\nin where the baroclinic mode is a high mode, rather than the first baroclinic mode considered by \cite{thomas_yamada2019}.

Two-dimensional models obtained from truncating three-dimensional equations to few modes have been used elsewhere in the literature, \textit{e.g.}, \cite{frierson_etal2004,benavides_alexakis2017}. We call this the modified Thomas \& Yamada (2019) model because \cite{thomas_yamada2019} treated only the particular case where $\Bu = 1$, relevant to the first mode of the internal tide rather than near-inertial waves. This difference can be seen by considering the nondimensional dispersion relation for inertia-gravity waves:

\beq
\omega^2 = f^2(1 + \Bu_w),
\eeq

\begin{figure}
\centering
\includegraphics[keepaspectratio=true,width=0.5\textwidth]{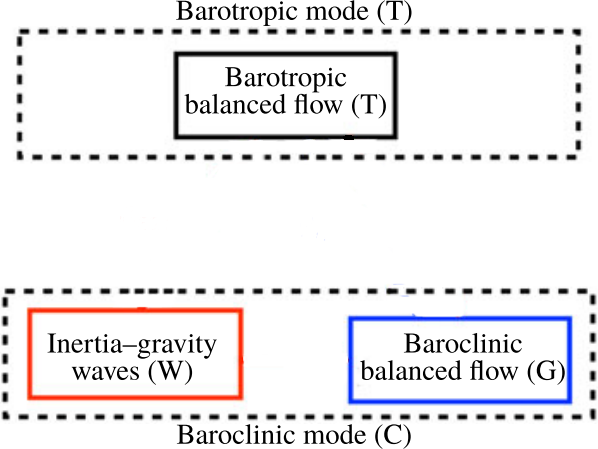}\\
\caption{Schematic showing the three components of the modified Thomas \& Yamada (2019) model: The barotropic mode ($T$-mode) consists only of purely geostrophically-balanced flow, while the baroclinic mode consists of geostrophically-balanced flow ($G$-mode) and unbalanced (near-inertial) wave motions ($W$-mode).}
\label{schemTY}
\end{figure}

\nin where $\omega$ is the wave frequency, $\Bu_w \equiv [Nk_h/(fk_z)]^2$ is the wave Burger number, $k_h \equiv \sqrt{k_x^2 + k_y^2}$ is the magnitude of the horizontal wavenumber vector and $k_z$ is the vertical component of the wavenumber vector. Since near-inertial waves have more energy content in high baroclinic modes, $\omega \approx f$, due to which $\Bu_w \ll 1$. This is the limit considered in this project. The dynamical components of the system described by \ref{ty11}-\ref{ty33} are represented schematically in Figure \ref{schemTY}. The barotropic mode ($T-$mode) contains only geostrophically-balanced energy, while the baroclinic mode contains both balanced energy ($G-$mode) and unbalanced inertia-gravity wave energy.

In order to further specialize the model to study the interactions between near-inertial waves and balanced flows, we supress the $G$-mode at every time step by inverting the linear baroclinic potential vorticity $q \equiv \zeta_C - p_C = \zeta_G - p_G$ and subtracting out the balanced velocity from the total baroclinic velocity vector $\bm{u_C}$. This is possible because only the balanced flow projects on $q$, since near-inertial waves have no linear potential vorticity.

\subsection{The coupled T-W model} \label{sec:TW}

\begin{figure}
\centering
\includegraphics[keepaspectratio=true,width=0.5\textwidth]{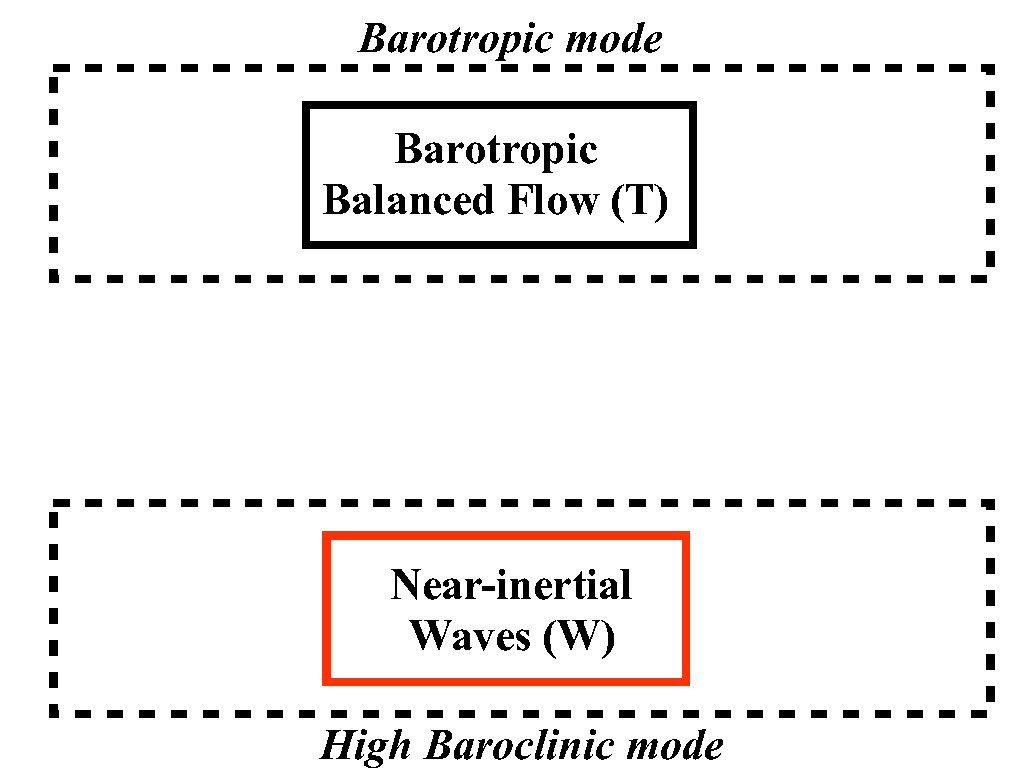}\\
\caption{Schematic showing the two components of the $T-W$ model: The barotropic mode ($T$-mode) consists only of purely geostrophically-balanced flow, while the high baroclinic mode consists only of unbalanced (near-inertial) wave motions.}
\label{schemTW}
\end{figure}

A more elegant and less artificial approach to isolate interactions between near-inertial waves and a balanced barotropic flow is to seek a simpler, two-component model (Figure \ref{schemTW}) where the prognostic variables are pure wave quantities. We start from the equations of motion (with the $C$ subscript dropped):

\beq
\partial_t\bm{u_T} + \bm{\hat{z}\times u_T} + \bm{\nabla}p_T = \bm{F^u_T},\label{eqmT}\\
\partial_t\bm{u} + \bm{\hat{z}\times u} + \Bu\bm{\nabla}p = \bm{F^u},\label{eqm1}\\
\partial_tp + \bm{\nabla\cdot u} = \bm{F^p},\label{eqm2}
\eeq

\nin where $\bm{u}$ and $\bm{u_T}$ are respectively the baroclinic wavy and barotropic geostrophically balanced velocities and

\beq
\bm{F^u_T} \defn -\Ro\big[\bm{u_T\cdot\grad u_T} + \bm{u\cdot\grad u} + (\divg\bm{u})\bm{u}\big],\\
\bm{F^u} \defn -\Ro\big(\bm{u_T\cdot\grad u} + \bm{u\cdot\grad u_T}\big),\\
\bm{F^p} \defn -\Ro\big(\bm{u_T\cdot\grad}p\big)
\eeq

\nin We can define a velocity potential $\phi$ and a streamfunction $\chi$ such that

\beq
u = \phi_x - \chi_y,\\
v = \phi_y + \chi_x,\\
u + iv = (\underbrace{\partial_x + i\partial_y}_{\defn \partial_s})(\underbrace{\phi + i\chi}_{\defn A}),
\eeq

\nin and

\beq
\phi = \phi_W,\\
\chi = \chi_G + \chi_W,\\
p = p_G + p_W,\\
\eeq

\nin where the subscripts $G$ and $W$ indicate baroclinic balanced (G-mode in Figure \ref{schemTY}) and wave (W-mode in Figure \ref{schemTY}) quantities, respectively. The G-mode satisfies

\beq
\zhat\times\bm{u_G} + \Bu\grad p_G = 0,\label{uG1}\\
\divg\bm{u_G} = 0,
\eeq

\nin while the W-mode satisfies

\beq
\partial_t\bm{u_W} + \zhat\times\bm{u_W} + \Bu\grad p_W = 0,\\
\partial_tp_W + \divg\bm{u_W} = 0.
\eeq

\nin Defining the Laplacian operator $\triangle \equiv \partial_x^2 + \partial_y^2$ and taking $\divg$\ref{uG1}, we have

\beq
-\zeta_G + \Bu\triangle p_G = 0 \Rightarrow \Bu\triangle p_G - \triangle\chi_G = 0 \Rightarrow \chi_G = \Bu p_G. \label{chiGBupG}
\eeq

\nin We form a conservation statement for the linear potential vorticity by setting $\Ro = 0$ in \ref{eqmT}, \ref{eqm1} and \ref{eqm2} and taking $\curl$\ref{eqm1} - \ref{eqm2}, defining $\zeta = \curl\bm{u}$:

\beq
\partial_t(\zeta - p) = 0 \Rightarrow \partial_t(\zeta_G - p_G) + \partial_t(\zeta_W - p_W) = 0. \label{pvq}
\eeq

\nin at this point we note that in the present model, inertia-gravity waves have no linear potential vorticity, and therefore the combination $\zeta_W - p_W$ is identically zero. This implies

\beq
p_W = \zeta_W \Rightarrow p_W = \triangle\chi_W. \label{pWchiW}
\eeq

\nin Back to \ref{pvq}, we have

\beq
\partial_t(\underbrace{\zeta_G - p_G}_{\equiv q_G}) = 0.
\eeq

\nin If we now make the choice $q_G = 0$ at $t = 0$, it follows that

\beq
p_G = \zeta_G \Rightarrow p_G = \triangle\chi_G. \label{pvqG}
\eeq

\nin Using \ref{chiGBupG} in \ref{pvqG}, it follows that

\beq
p_G = \Bu\triangle p_G. \label{pGBupG}
\eeq

\nin The only way that \ref{pGBupG} can be satisfied is if $p_G = 0$ for all $t$, from which it follows that $\chi_G = 0$, $p = p_W$ and $\chi = \chi_W$. The choice of zero baroclinic potential vorticity $q_G$ thus eliminates the G-mode, resulting in the two-component system represented in Figure \ref{schemTW}.

The next step is to obtain evolution equations for $\phi$ and $\chi$. Taking $\divg$\ref{eqm1}, $\curl$\ref{eqm1} and $\triangle$\ref{eqm2} yields, respectively,

\beq
\partial_t(\overbrace{\divg\bm{u}}^{\triangle\phi}) - \triangle\chi + \Bu\triangle p = \divg\bm{F^u},\label{eqm3}\\
\partial_t(\underbrace{\curl\bm{u}}_{\triangle\chi}) + \triangle\phi = \curl\bm{F^u}\label{eqm4},\\
\partial_t(\triangle p) + \triangle^2\phi = \triangle\bm{F^p}.\label{eqm5}
\eeq

\nin Taking \ref{eqm4} - $\Bu\times$\ref{eqm5},

\beq
\partial_t\triangle\big(\chi - \Bu p\big) = - \triangle(1 - \Bu\triangle)\phi + \curl\bm{F^u} - \Bu\triangle\bm{F^p}. \label{chi1}
\eeq

\nin We then note that

\beq
\chi - \Bu p = \underbrace{\chi_G - \Bu p_G}_{=0} + \chi_W - \Bu p_W = \chi_W - \Bu p_W, \label{chiBup}
\eeq

\nin where the last equality in \ref{chiBup} follows from \ref{chiGBupG}. Using \ref{pWchiW} in \ref{chi1}:

\beq
\partial_t\triangle\big(\chi - \Bu\triangle\chi\big) = - \triangle(1 - \Bu\triangle)\phi + \curl\bm{F^u} - \Bu\triangle\bm{F^p}.
\eeq

\nin or

\beq
\boxed{
\partial_t\chi = \triangle^{-1}\big(1 - \Bu\triangle\big)^{-1}\curl\bm{F^u} - \Bu\big(1 - \Bu\triangle\big)^{-1}\bm{F^p} - \phi.
}
\label{eqtpsi}
\eeq

\nin By using \ref{pWchiW} in \ref{eqm3}, we obtain

\beq\label{eqtphi}
\partial_t(\triangle\phi) = \triangle(1 - \Bu\triangle)\chi + \divg\bm{F^u},
\eeq

\nin or

\beq\label{eqtphi2}
\boxed{
\partial_t\phi = (1 - \Bu\triangle)\chi + \triangle^{-1}(\divg\bm{F^u}).
}
\eeq

\nin Taking $\curl$\ref{eqmT} and noting that $\divg \bm{u_T} = 0$ gives the evolution equation for the T-mode's vorticity $\zeta_T = \triangle\Psi$, where $\Psi$ is the T-mode's streamfunction:

\beq\label{eqPsi}
\boxed{
\partial_t\zeta_T = \curl\bm{F^u_T}.
}
\eeq

\nin We now seek an evolution equation for the purely inertial wave mode (the zeroth wavenumber wave). This mode's velocity is spatially uniform across the domain, and therefore can be isolated by taking the spatial average (denoted by $\langle\bullet\rangle$) of the momentum equation \ref{eqm1}:

\beq
\langle\bm{u}_t\rangle + \bm{\hat{z}\times \langle u\rangle} + \langle\Bu\bm{\nabla}p\rangle = \langle\bm{F^u}\rangle \label{savgu}
\eeq

\nin Defining $A_0 \equiv \langle u\rangle + i\langle v\rangle = u_0 + iv_0$, the $\xhat$ and $\yhat$ components of \ref{savgu} are, respectively,

\beq
\partial_tu_0 - v_0 = \Re\langle\bm{F^u}\rangle,\label{inx} \\
\partial_tv_0 + u_0 = \Im\langle\bm{F^u}\rangle \label{iny}
\eeq

\nin Taking \ref{inx} + $i\times$\ref{iny} yields

\beq\label{eqA0}
\boxed{
\partial_tA_0 = -iA_0 + \Re\langle\bm{F^u}\rangle + i\Im\langle\bm{F^u}\rangle.
}
\eeq

Equations \ref{eqtpsi}, \ref{eqtphi2}, \ref{eqPsi} and \ref{eqA0} form a set of evolution equations for the barotropic vorticity $\zeta_T$ (T-mode), the near-inertial wave amplitude $A \equiv \phi + i\chi$ (W-mode) and the pure inertial oscillation's amplitude $A_0$.

\subsubsection{Energetics of the coupled T-W model}

We may obtain equations for the kinetic energy of the divergent and rotational parts of the wave velocity by taking $\phi\times$\ref{eqtphi} and $\chi\times\triangle$\ref{eqtpsi}, respectively, and integrating over the domain. The result is (recalling from \ref{pWchiW} that $p = \triangle\chi$),

\beq\label{eqphiphi}
\frac{\partial}{\partial t}\frac{1}{2}\langle|\grad\phi|^2\rangle = \langle\grad\phi\cdot\grad\chi\rangle - \Bu\langle\grad\phi\cdot\grad(\triangle\chi)\rangle + \langle\bm{F^u}\cdot\grad\phi\rangle,
\eeq

\nin and

\beq\label{eqchichi}
\frac{\partial}{\partial t}\frac{1}{2}\langle|\grad\chi|^2\rangle = - \langle\grad\phi\cdot\grad\chi\rangle - \langle\chi(1 - \Bu\triangle)^{-1}\big(\curl\bm{F^u} - \Bu\triangle\bm{F^p}\big)\rangle.
\eeq

\nin The first term on the right-hand sides of \ref{eqphiphi} and \ref{eqchichi} appears with opposite signs in both equations and can therefore be interpreted as a conversion term that represents the kinetic energy transfers between the rotational and divergent parts of the wave field.

\nin Recalling again that $p = \triangle\chi$, we can also form a potential energy equation by taking $\triangle\chi\times\triangle$\ref{eqtpsi} and integrating over the domain to obtain

\beq\label{EWPE}
\frac{\partial}{\partial t}\frac{1}{2}\langle(\triangle\chi)^2\rangle = - \langle\triangle\chi\triangle\phi\rangle + \langle\triangle\chi(1 - \Bu\triangle)^{-1}\big(\curl\bm{F^u} - \Bu\triangle\bm{F^p}\big)\rangle.
\eeq

\nin The pure inertial mode's energy equation can be obtained by taking $A_0^\star\times$\ref{eqA0}, adding the entire expression's complex conjugate and dividing the result by two (where the star superscript indicates the complex conjugate):

\beq\label{EA0}
\frac{\partial}{\partial t}\frac{1}{2}|A_0|^2 = \Re\langle\bm{F^u}\rangle\Re A_0 + \Im\langle\bm{F^u}\rangle\Im A_0.
\eeq

\nin Finally, the T-mode's energy equation can be obtained by taking $-\Psi\times$\ref{eqPsi} and integrating over the domain.

\beq\label{ET}
\partial_t\frac{1}{2}\langle\Psi\zeta_T\rangle = \langle\Psi\curl\bm{F^u_T}\rangle.
\eeq

\nin It can be verified numerically that the system conserves total energy, that is, (\ref{eqphiphi}) + (\ref{eqchichi}) + $\Bu\times$(\ref{EWPE}) + (\ref{EA0}) + (\ref{ET}) = 0:

\beq
\frac{\partial}{\partial_t}\frac{1}{2}\iint |\grad\phi|^2 + |\grad\chi|^2 + \Bu(\triangle\chi)^2 + |A_0|^2 -\Psi\zeta_T \, dA = 0.
\eeq

\section{Parameter sweep with the linearized modified Thomas \& Yamada (2019) model}

\begin{figure}
\centering
\includegraphics[keepaspectratio=true,width=0.5\textwidth]{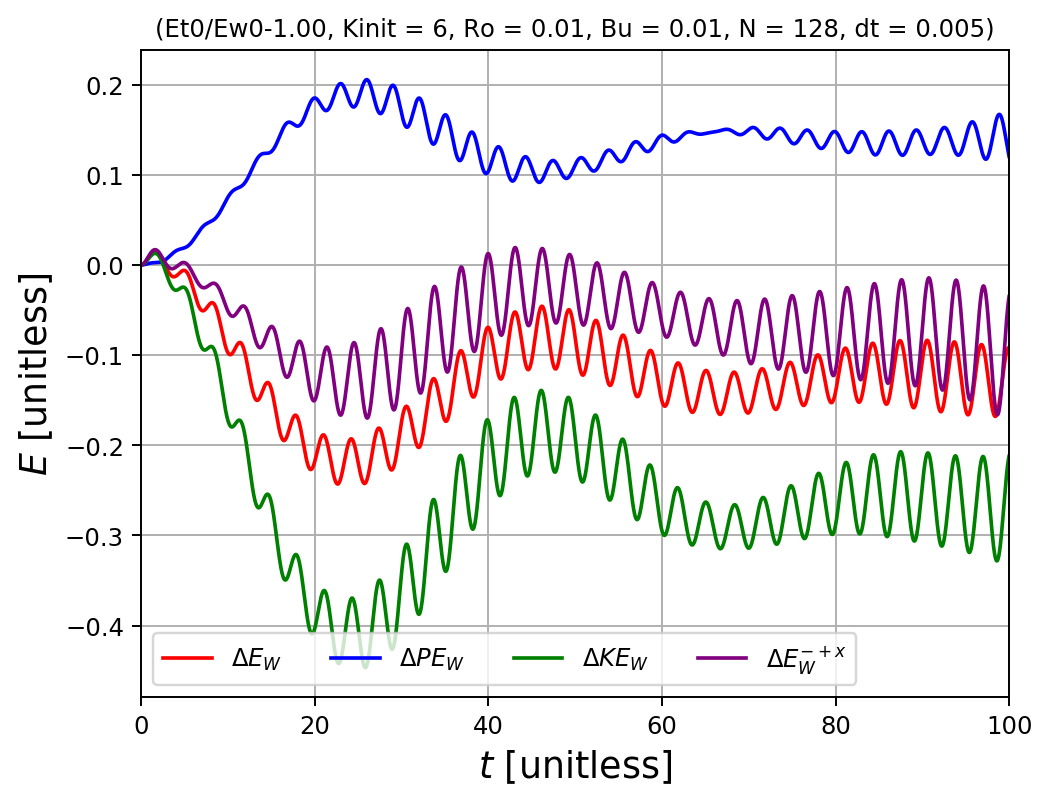}\includegraphics[keepaspectratio=true,width=0.5\textwidth]{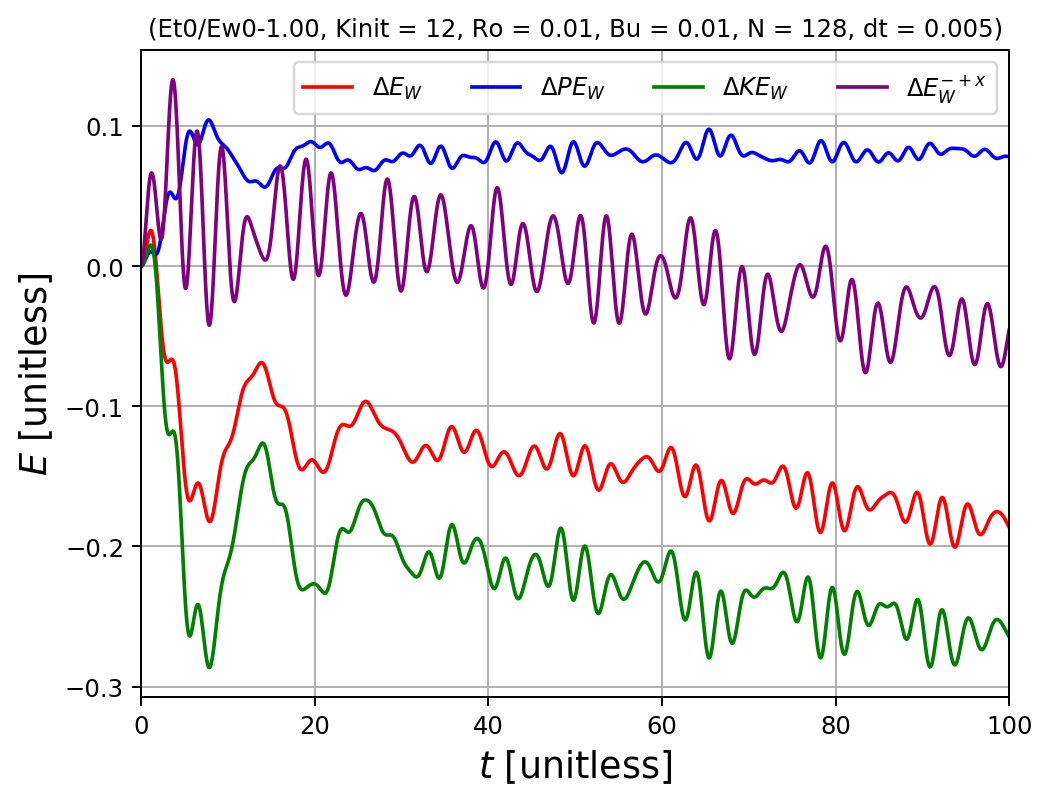}\\
\includegraphics[keepaspectratio=true,width=0.5\textwidth]{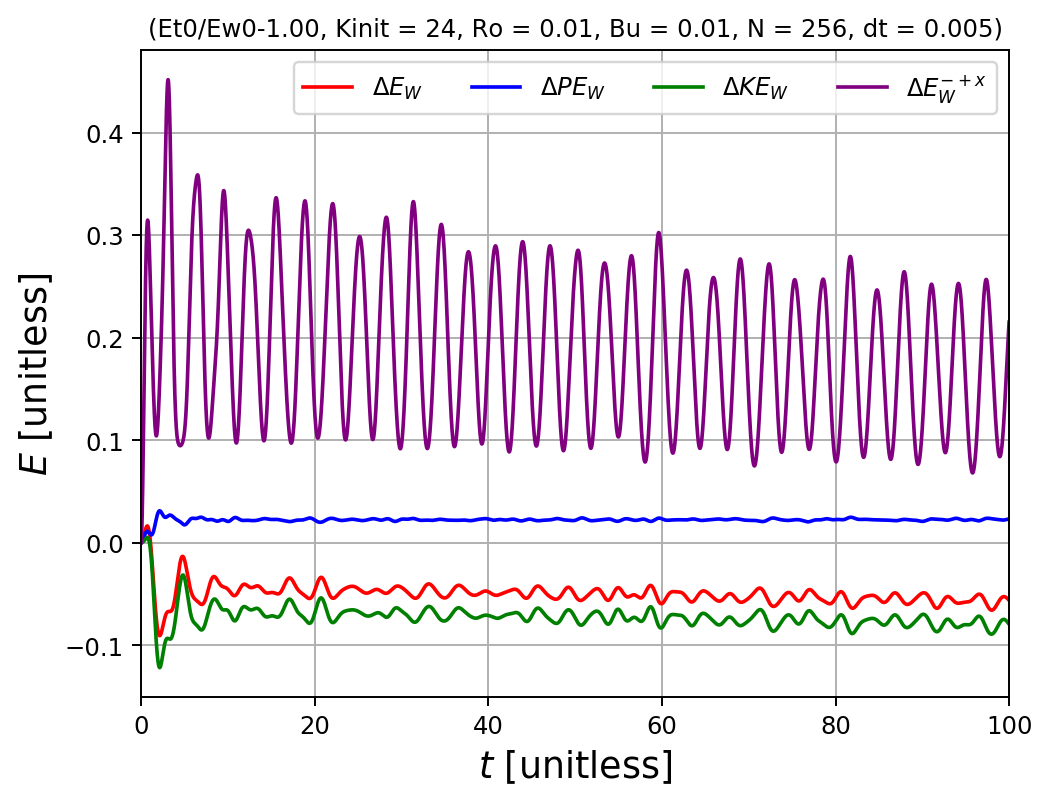}
\caption{Wave energy changes for 6 (top left) 12 (top right) and 24 (bottom) initial wavenumbers in the system evolving according to Equations \ref{ty11}-\ref{ty33}. Note that the wave potential energy gain is offset by the wave kinetic energy loss, causing the total wave energy to decrease.}
\label{simKinit}
\end{figure}

Next, we explore the sensitivity of the energy changes to different barotropic flows in the linearized version of \ref{ty11}-\ref{ty33} (linearized about a steady barotropic balanced flow $\bm{U} = \bm{\hat{x}}U + \bm{\hat{y}}V$). We solve \ref{ty11}-\ref{ty33} using a standard pseudo-spectral code based on \cite{thomas_etal2017}. Figure \ref{simKinit} compares the evolution of wave kinetic, potential and total energies for simulations with barotropic balanced flows with randomized phase and increasing number of initial wavenumbers $K_i$.

The wave amplitude can be further approximately decomposed into clockwise and counter-clockwise motions as follows:

\beq
\U = A^-e^{-it} + A^+e^{it}
\eeq

\nin So that the kinetic energy is

\beq
\frac{1}{2}\iint\overline{\U}\U \, dx\, dy = \frac{1}{2}\iint |A^-|^2 + |A^+|^2 + \overline{A^-}A^+e^{+2it} + \overline{A^+}A^-e^{-2it}\, dx\, dy
\eeq

\nin and the potential energy is (using the fact that $p^+ = \overline{p^-}$)

\beq
\frac{1}{2}\iint\overline{p}p \, dx\, dy = \frac{1}{2}\iint 2|p^-|^2 + \overline{p^-}\,\overline{p^-}e^{+2it} + p^-p^-e^{-2it}\, dx\, dy
\eeq

\begin{figure}
\centering
\includegraphics[keepaspectratio=true,width=0.5\textwidth]{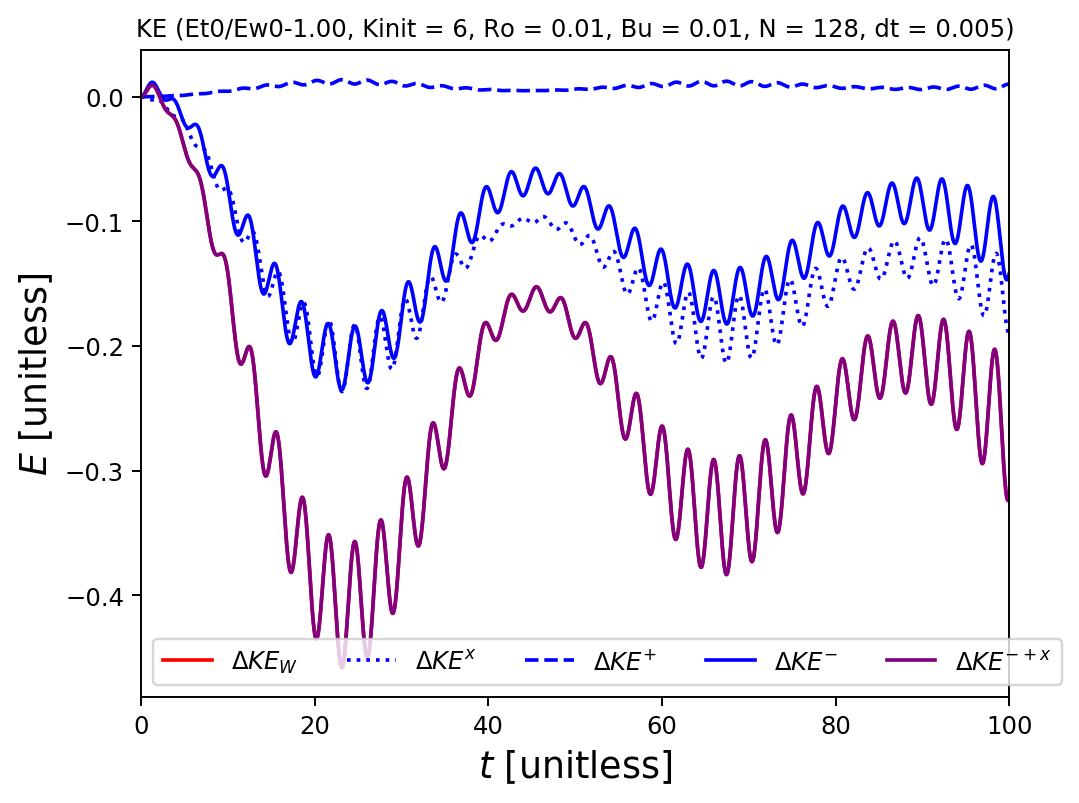}\includegraphics[keepaspectratio=true,width=0.5\textwidth]{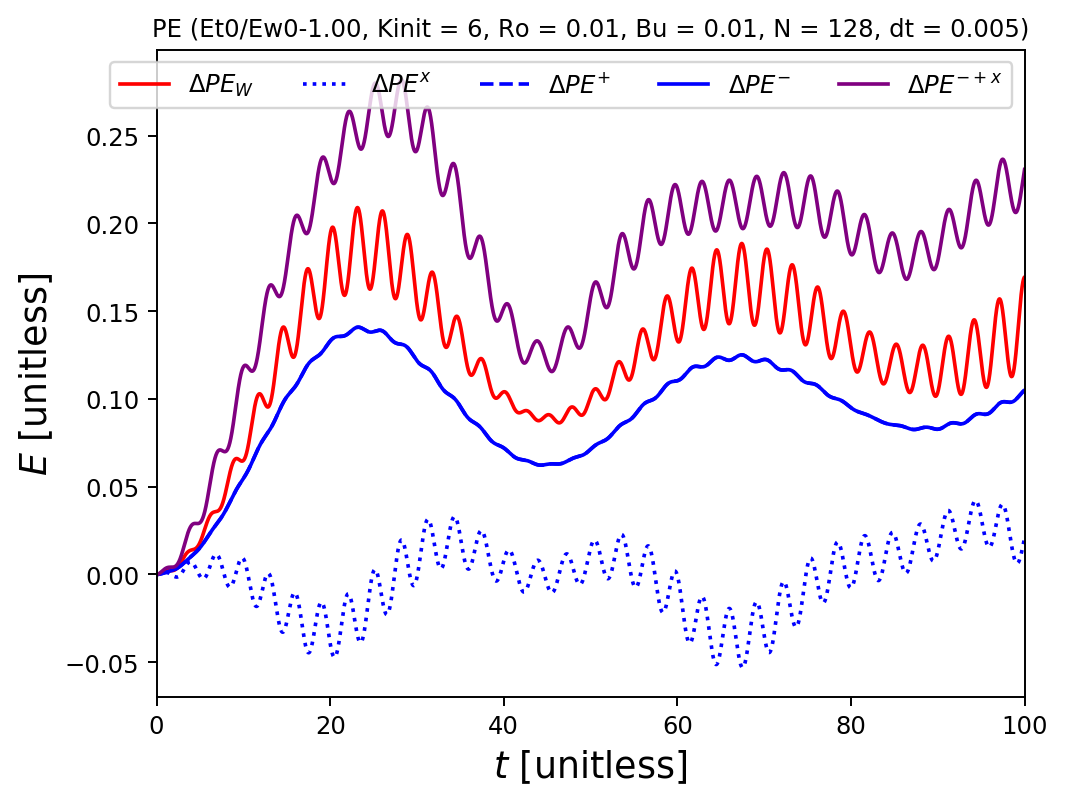}\\
\includegraphics[keepaspectratio=true,width=0.5\textwidth]{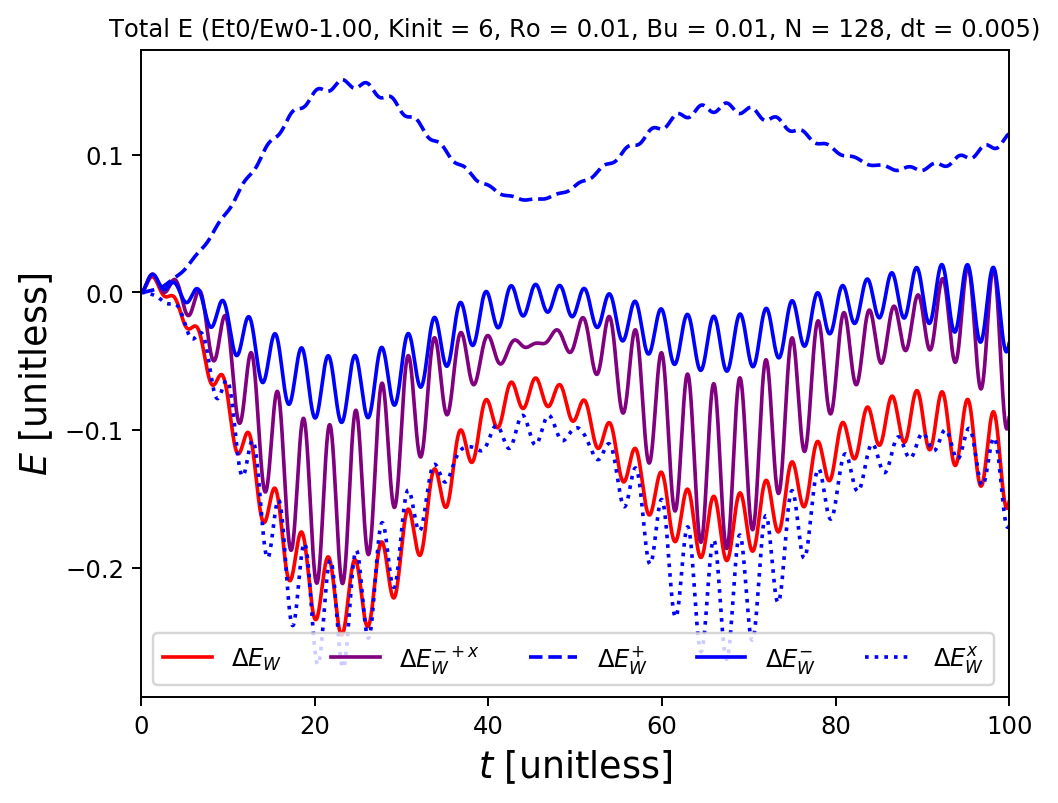}\\
\caption{Kinetic (top left), potential (top right) and total (bottom) wave energy changes in the system evolving according to Equations \ref{ty11}-\ref{ty33} for the approximate decomposition in $+$, $-$ and $x$ (cross) terms. Note that the cross terms are not negligible, indicating that this simplified decomposition is non-orthogonal.}
\label{simxterms}
\end{figure}

Figure \ref{simxterms} compares the evolution of the wave energy terms associated with positive (proportional to $A^+$) negative (proportional to $A^-$) and mixed (proportional to $\overline{A^+}A^-$ + c.c.) amplitudes. It can be seen that the cross component has magnitude comparable to the $+$ and $-$ components, indicating that this decomposition is non-orthogonal, contrary to the orthogonal decompositions used in \tit{e.g.}, \cite{remmel_smith2009,thomas_yamada2019}.

\section{Helmholtz decomposition of reduced model solutions}

In this section we briefly compare the energy partitioning into rotational (balanced, non-divergent) and divergent (unbalanced, irrotational) motions in the linearized version of \ref{ty11}-\ref{ty33} with the partitioning in the simpler Young and Ben Jelloul (YBJ, \cite{young_benjelloul1997}) model. The total velocity field can be decomposed into a velocity potential $\phi$ (irrotational) and a streamfunction $\psi$ (non-divergent) according to

\beq
\psi = \triangle^{-1}\big(v_x - u_y\big),\\
\phi = \triangle^{-1}\big(u_x + v_y\big),\\
u + iv = (\partial_x + i\partial_y)(\phi + i\psi).
\eeq

\begin{figure}
\centering
\includegraphics[keepaspectratio=true,width=0.5\textwidth]{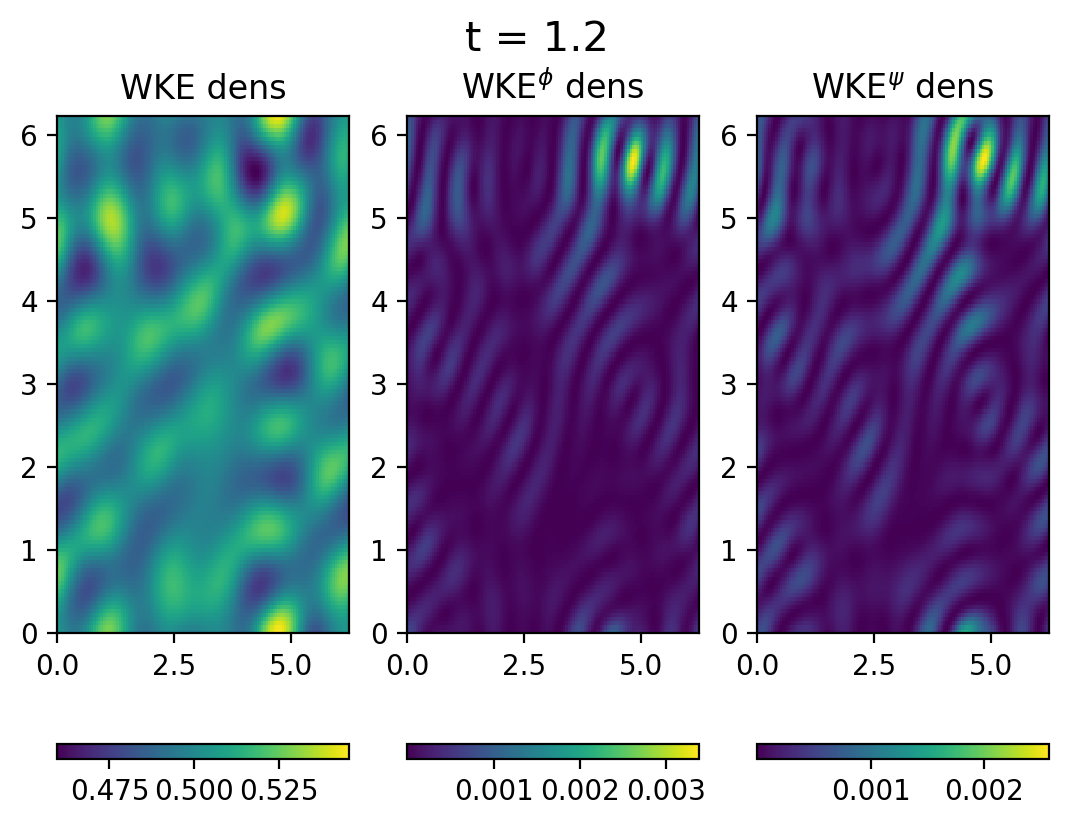}\includegraphics[keepaspectratio=true,width=0.5\textwidth]{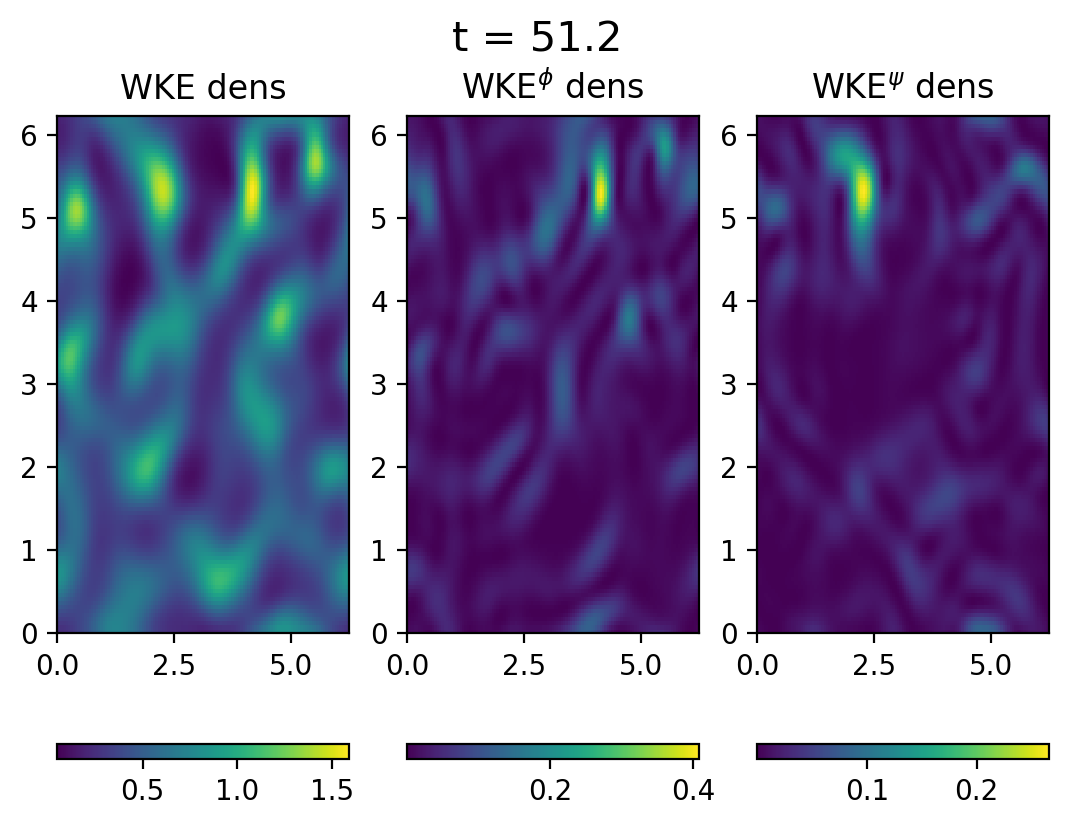}\\
\includegraphics[keepaspectratio=true,width=0.5\textwidth]{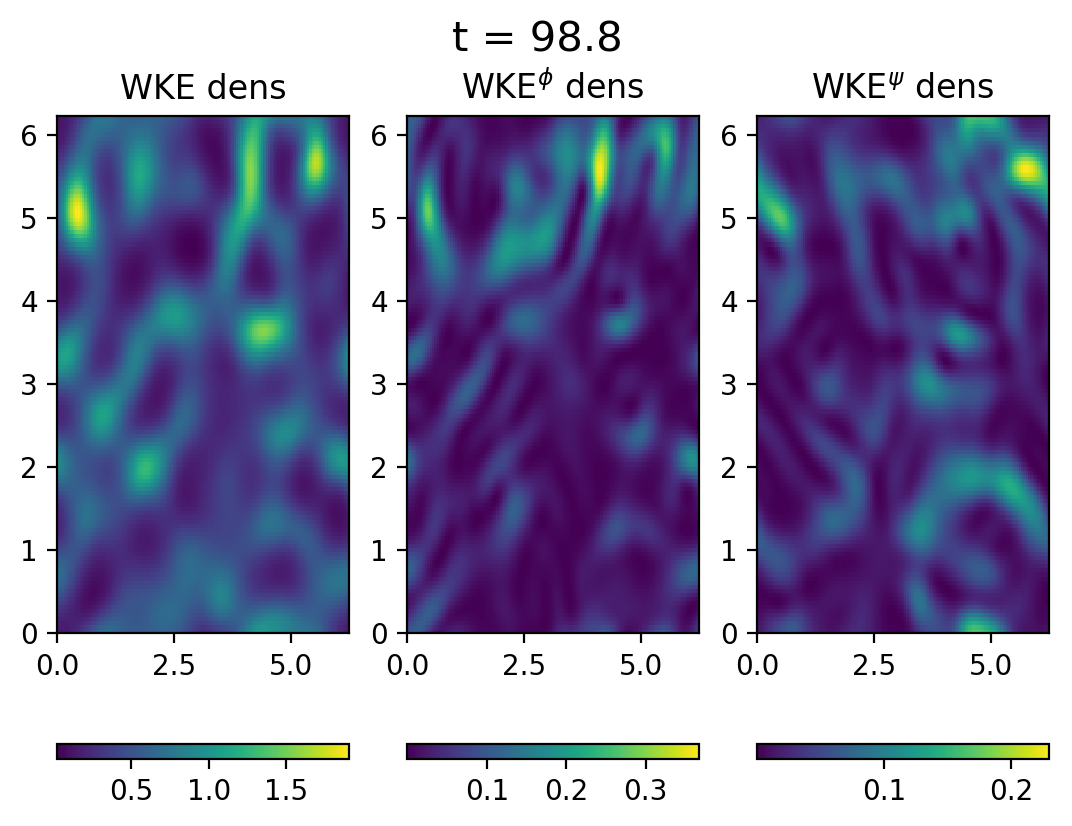}\\
\caption{Kinetic energy density snapshots of a linearized \ref{ty11}-\ref{ty33} solution decomposed into rotational ($\psi$) and divergent ($\phi$) parts.}
\label{Helm_TSB_xy}
\end{figure}

\begin{figure}
\centering
\includegraphics[width=0.5\linewidth,keepaspectratio=true]{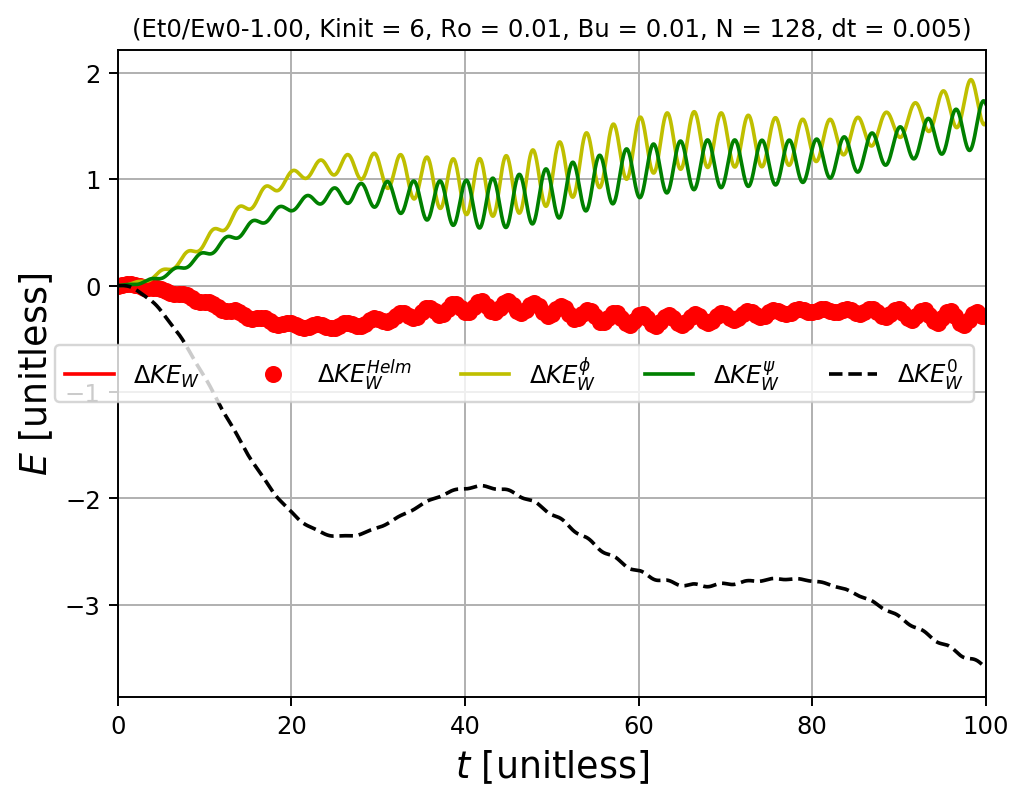}\includegraphics[width=0.5\linewidth,keepaspectratio=true]{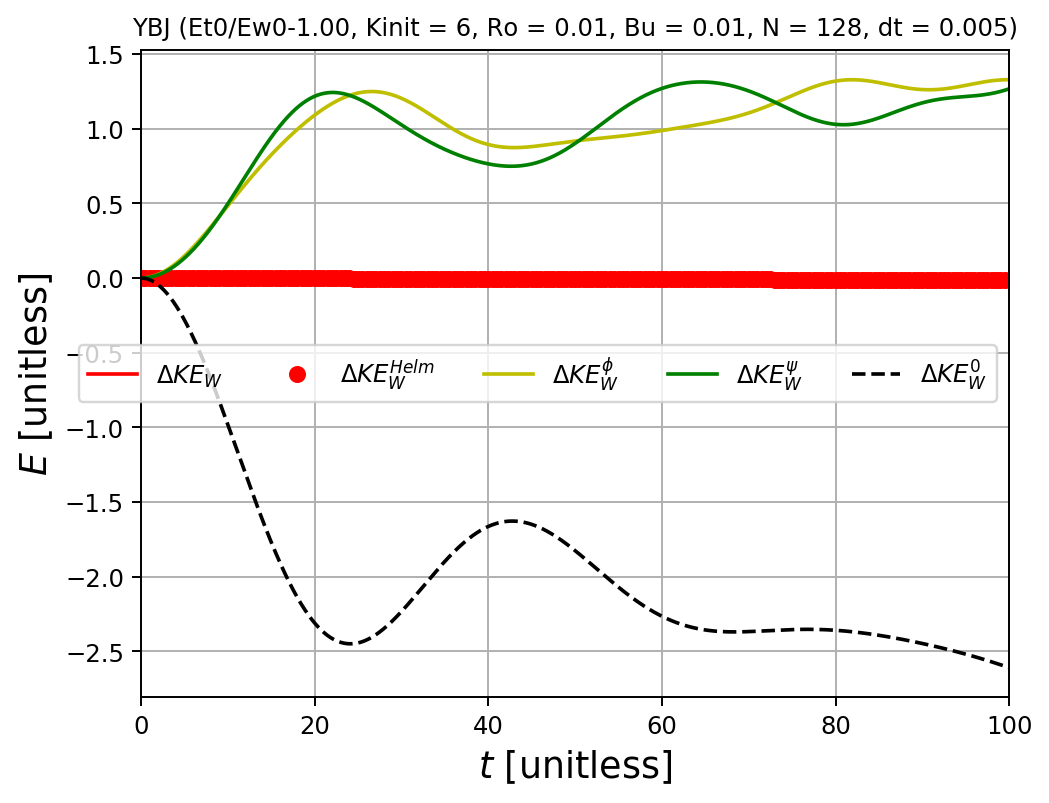}
\caption{Energy changes for the linearized \ref{ty11}-\ref{ty33} system (left) and the YBJ system (right). The kinetic energy is decomposed into rotational ($\psi$, green) and divergent ($\phi$, yellow) parts.}
\label{Helm_TSB-YBJ}
\end{figure}

\nin Figure \ref{Helm_TSB_xy} shows the spatial distribution of the wave kinetic energy density in a simulation of the linearized \ref{ty11}-\ref{ty33} system. Figure \ref{Helm_TSB-YBJ} shows the energy evolution in different reservoirs, which is similar in both systems: The purely inertial mode (inertial oscillations evolving according to \ref{eqA0}, dashed black lines) loses kinetic energy while the near-inertial modes gain kinetic energy. This energy gain is approximately equipartitioned between rotational and divergent motions. It can also be seen that the wave kinetic energy decreases in the linearized \ref{ty11}-\ref{ty33} solution, while it stays constant in the YBJ solution, as predicted by one of its conservation laws \cite{young_benjelloul1997}. The fact that the YBJ system conserves wave kinetic energy is one of its limitations.

\section{Wave-balanced flow interaction: Case studies}

In this section we aim to gain some physical intuition on the wave-balanced flow interaction by analyzing a set of initial value problems with different barotropic flows as initial conditions. Specifically, we seek answers to the following questions:

\begin{enumerate}
\item{How do different balanced flows couple with the near-inertial wave field?}
\item{what is the direction of the energy transfers, \tit{i.e.}, from balanced flow to waves or from waves to balanced flow?}
\end{enumerate}

The system simulated is \ref{ty11}-\ref{ty33}, where the barotropic flow evolves according to Equation \ref{ty11} and therefore has a two-way coupling with the near-inertial waves. The balanced part of the baroclinic mode is removed by inverting the potential vorticity. Although the two-component $T$-$W$ model has not been numerically implemented yet, it is expected to give similar results.

\begin{figure}
\centering
\includegraphics[keepaspectratio=true,width=0.5\textwidth]{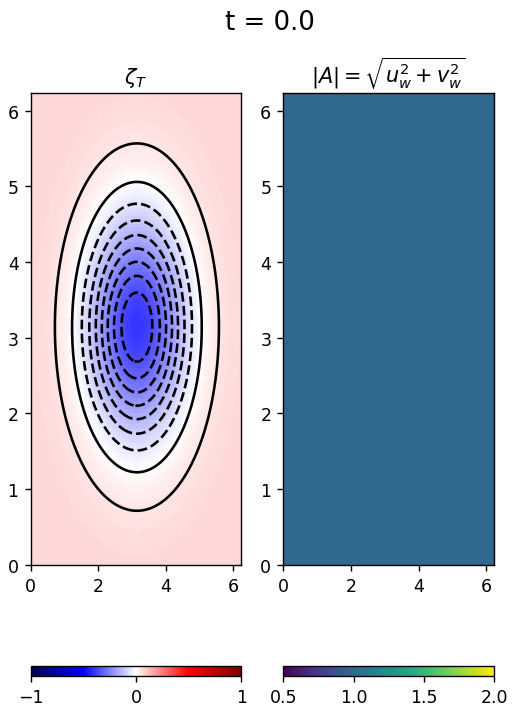}\includegraphics[keepaspectratio=true,width=0.5\textwidth]{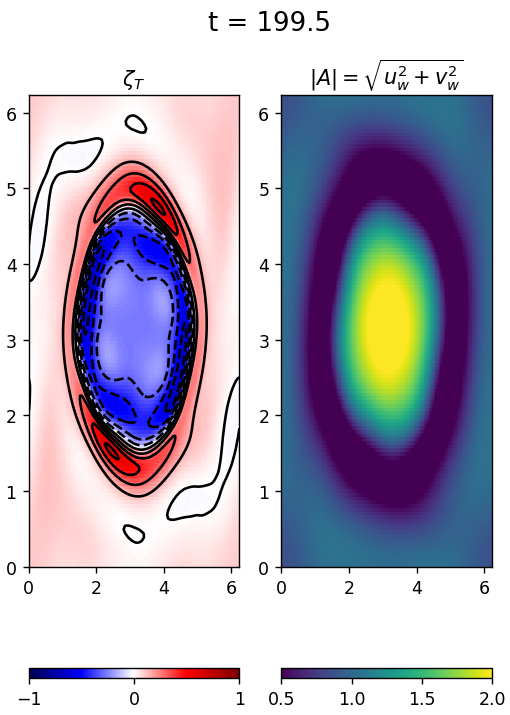}\\
\includegraphics[keepaspectratio=true,width=0.7\textwidth]{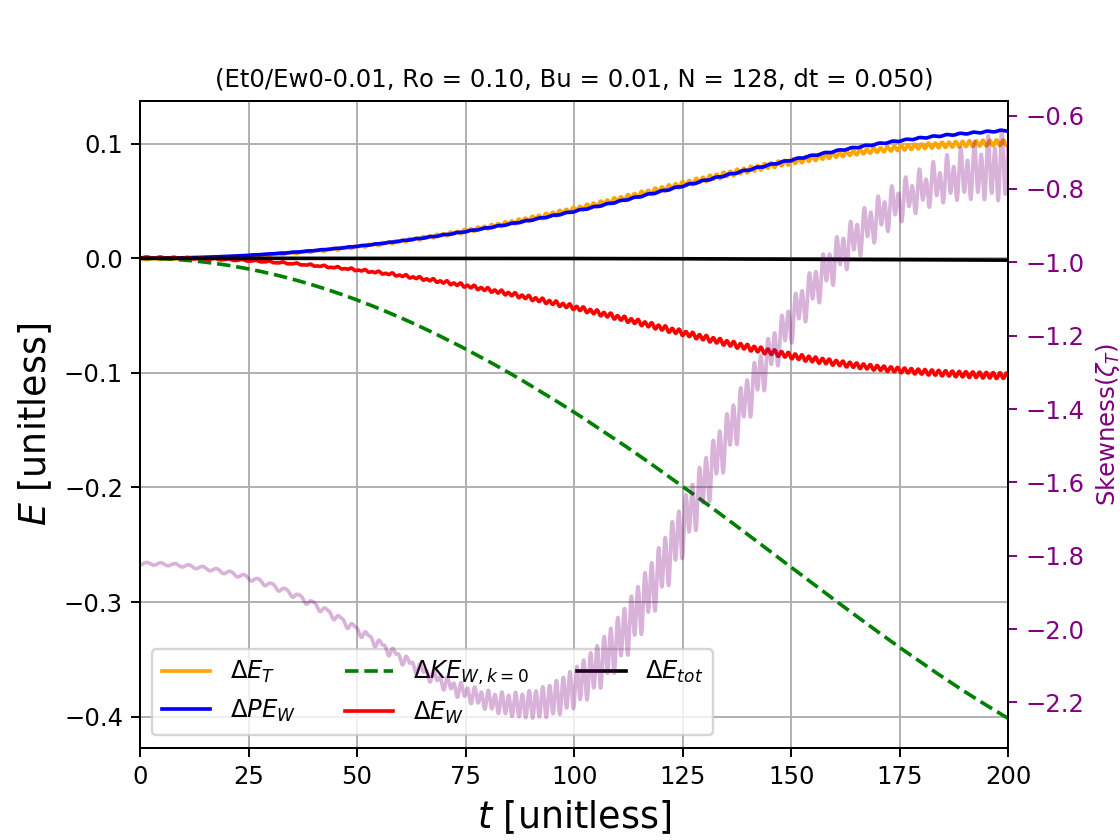}
\vspace*{-0.25cm}
\caption{Coupled T-W wave energy changes (lower panel) and snapshots of the barotropic vorticity $\zeta_T$ and the wave amplitude $|A|$ (upper panels), initialized with a barotropic anticyclone. $\Delta E_T$, $\Delta E_\text{tot}$, $\Delta PE_W$, $\Delta KE_{W,k=0}$ and $\Delta E_W$ are the balanced barotropic energy, the total (wave + balanced) energy, the wave potential energy, the wave kinetic energy in the purely inertial mode ($k=0$) and the total wave energy, respectively. The purple line in the lower panel is the instantaneous skewness of the barotropic vorticity. $E_{t0}/E_{w0}$ is the initial balanced-to-wave energy ratio, and $N$ and $dt$ are respectively the number of Fourier modes and the time step in the simulation.}
\label{simTWanticyclone}
\end{figure}

We begin with an initial barotropic flow of a simple Gaussian anticyclone (Figure \ref{simTWanticyclone}) in the presence of a spatially uniform inertial oscillation. As is well known in the literature (\tit{e.g.}, \cite{rocha_etal2018}), wave kinetic energy density gets trapped inside anticyclonic vortices, as observed in this experiment (top-right panels in Figure \ref{simTWanticyclone}). The total balanced energy increases at the expense of the total wave energy, and the skewness of the barotropic vorticity changes from negative to slightly positive by $t = 200$ (bottom panel of Figure \ref{simTWanticyclone}), indicating a change in the predominance of anticyclones (negative vorticity) to cyclones (positive vorticity).

\begin{figure}
\centering
\includegraphics[keepaspectratio=true,width=0.5\textwidth]{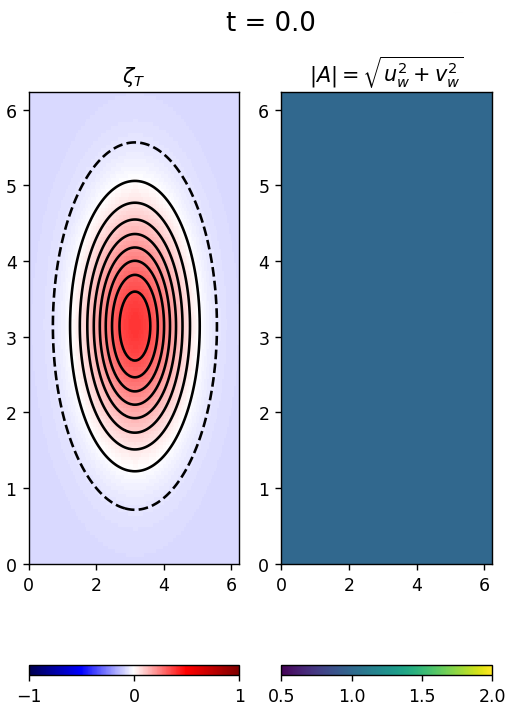}\includegraphics[keepaspectratio=true,width=0.5\textwidth]{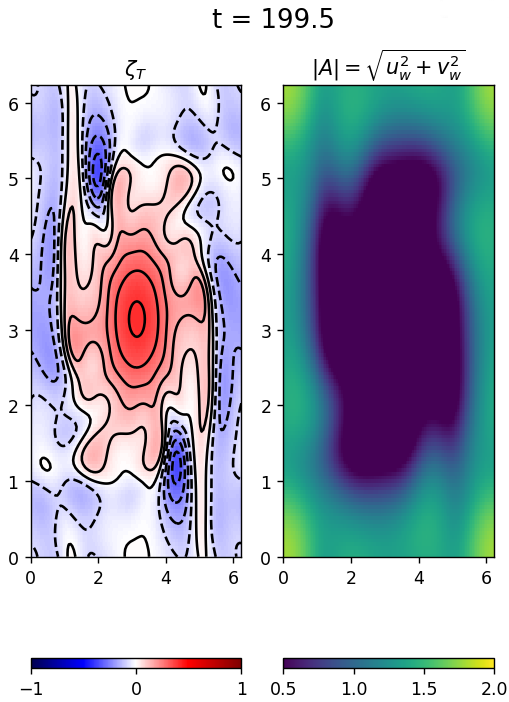}\\
\includegraphics[keepaspectratio=true,width=0.7\textwidth]{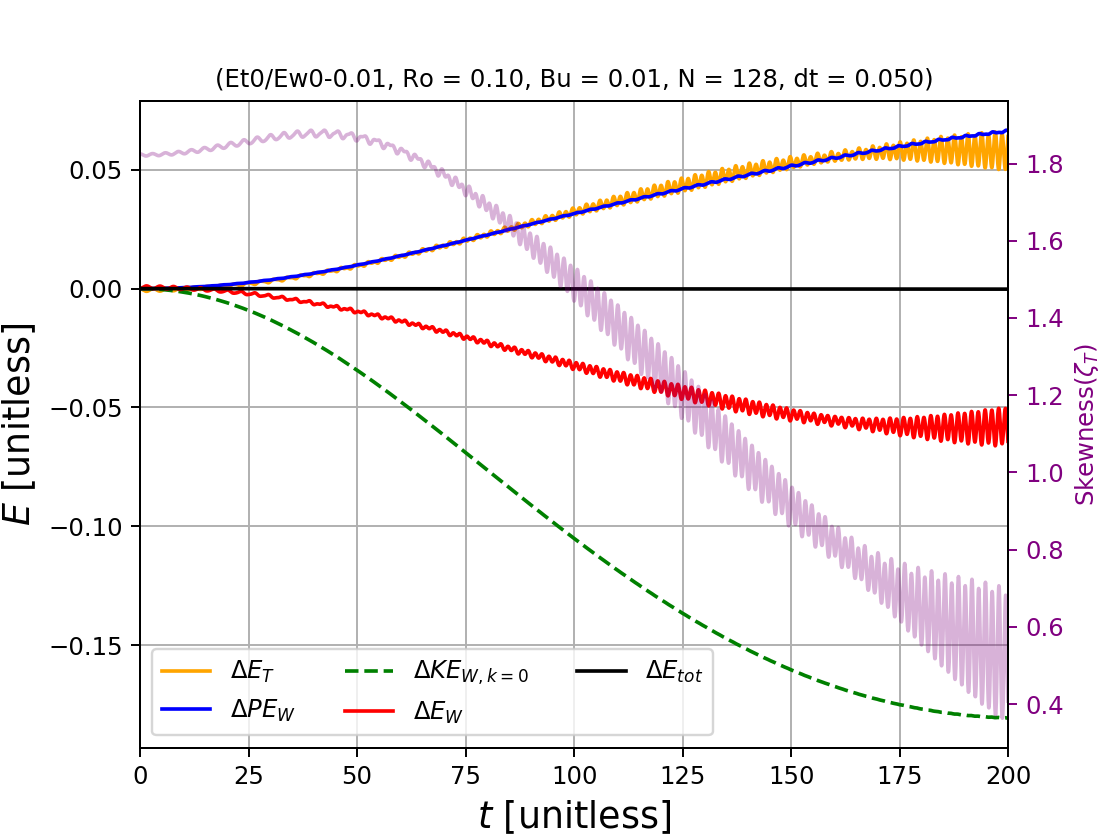}
\caption{Same as Figure \ref{simTWanticyclone} but initialized with a barotropic cyclone.}
\label{simTWcyclone}
\end{figure}

When the initial barotropic flow is a cyclone superimposed on a spatially uniform inertial oscillation, wave kinetic energy density is repelled from the core of the vortex (top-right panels of Figure \ref{simTWcyclone}), contrary to the anticyclonic case described in the previous paragraph. The skewness changes from positive to negative, also in contrast with the anticyclonic case. However, the energy changes of the cyclonic case are qualitatively similar to the anticyclonic case (bottom panel of Figure \ref{simTWcyclone}).

\begin{figure}
\centering
\includegraphics[keepaspectratio=true,width=0.5\textwidth]{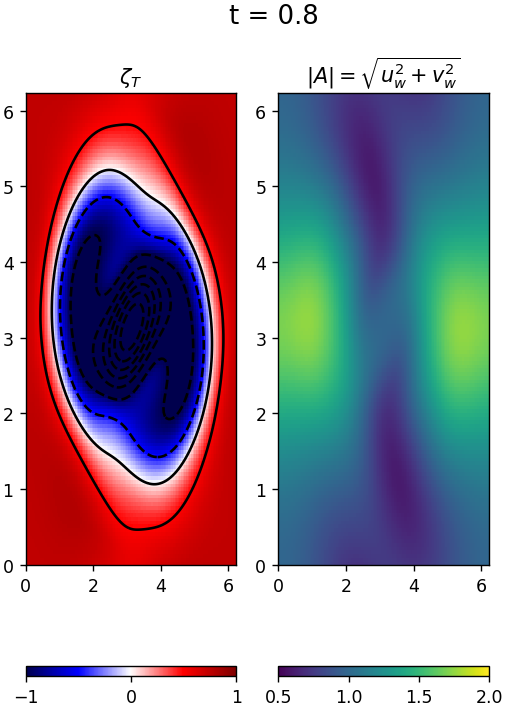}\includegraphics[keepaspectratio=true,width=0.5\textwidth]{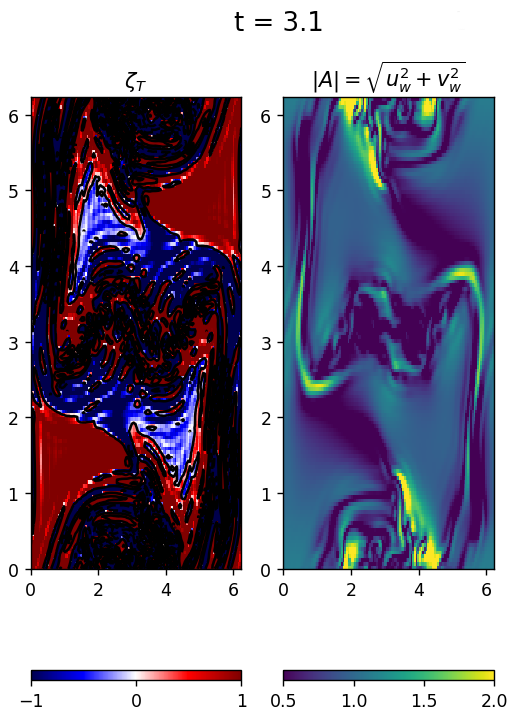}\\
\includegraphics[keepaspectratio=true,width=0.7\textwidth]{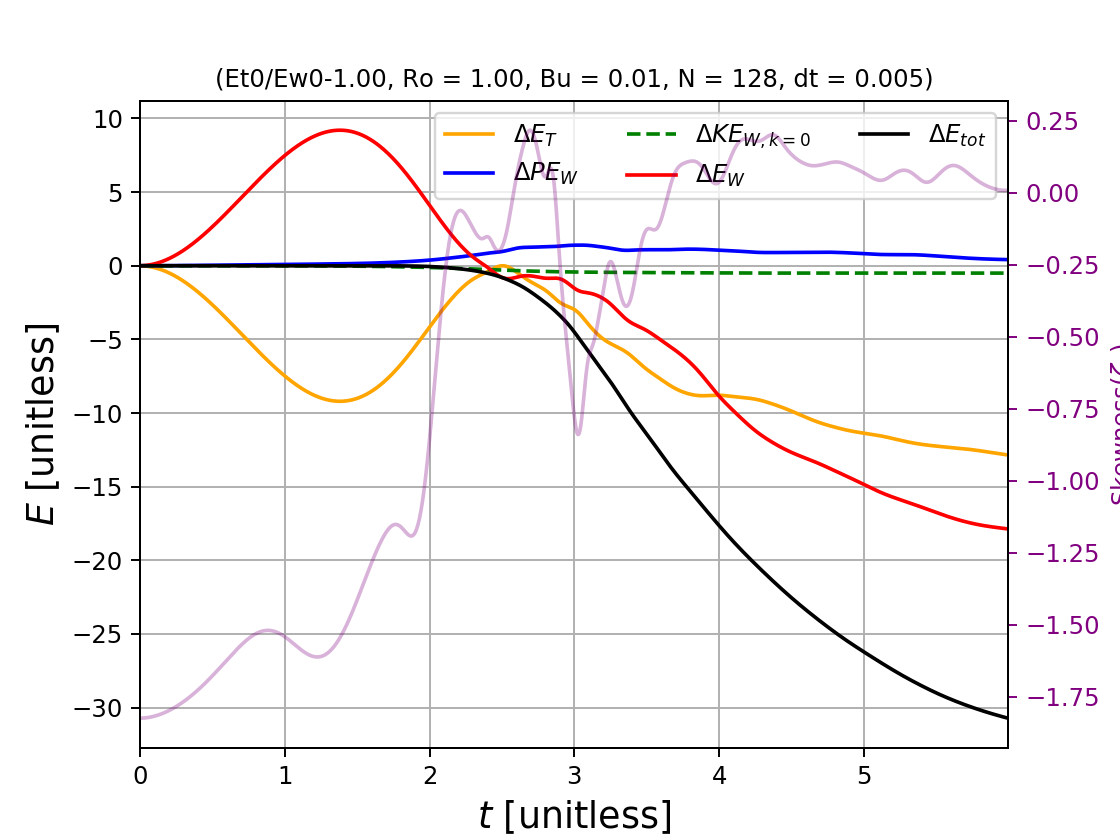}
\caption{Same as Figure \ref{simTWanticyclone}, but initialized with a barotropic anticyclone with $\Ro = 1$.}
\label{simTWanticycloneRo1}
\end{figure}

Does this energy pathway change direction as $\Ro \to 1$? Figure \ref{simTWanticycloneRo1} shows results of a simulation identical to that in Figure \ref{simTWanticyclone}, except for the Rossby number, which is set to 1. Numerical instability sets in very early on in the simulation, and total energy is no longer conserved after $t \approx 2.5$. However, if not an initial transient or a numerical artifact, the behavior seen at $t<2.5$ could suggest that the energy exchange changes direction, with waves now extracting energy from the balanced flow. This would imply that it is possible to reproduce the behavior of fully three-dimensional, non-hydrostatic Boussinesq simulations at $\Ro \sim 1$ (\tit{e.g.}, \cite{barkan_etal2017}) with this simple two-dimensional model.

\begin{figure}
\centering
\includegraphics[keepaspectratio=true,width=0.5\textwidth]{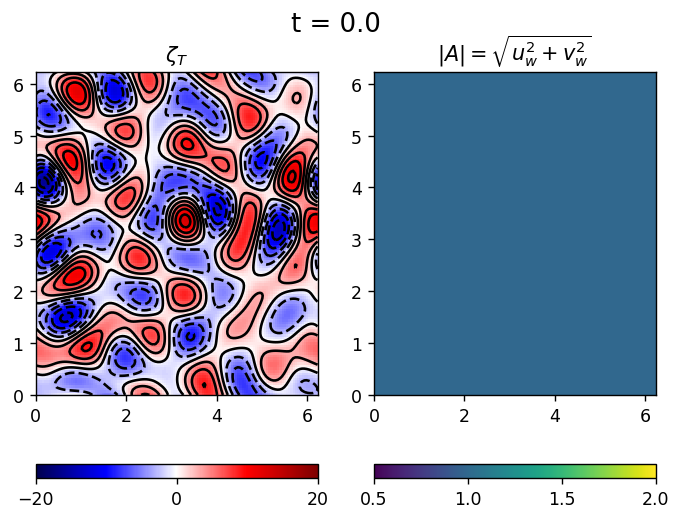}\includegraphics[keepaspectratio=true,width=0.5\textwidth]{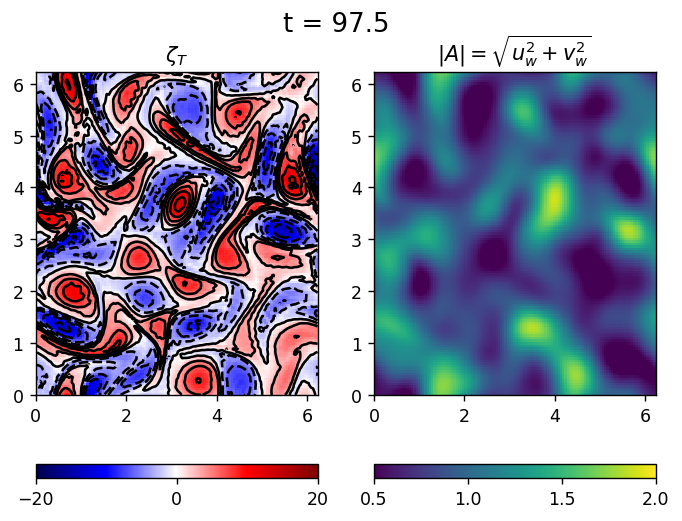}\\
\includegraphics[keepaspectratio=true,width=0.7\textwidth]{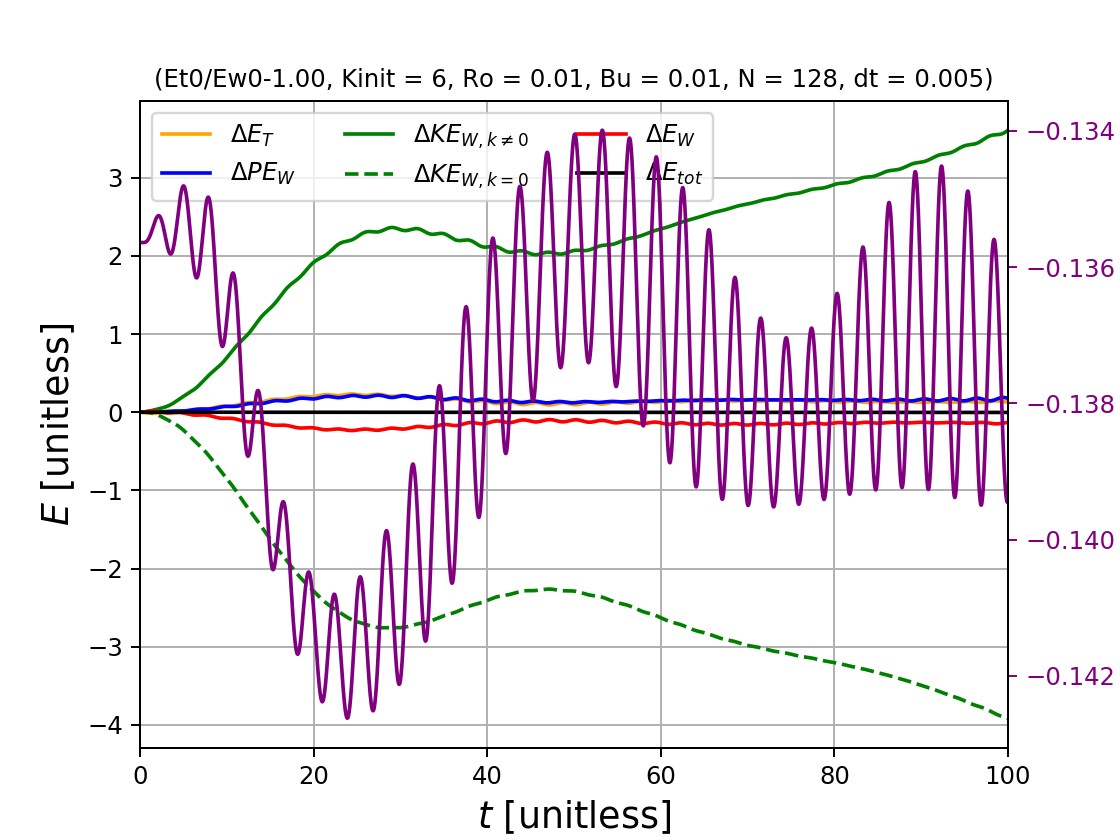}
\caption{Same as Figure \ref{simTWanticyclone}, but initialized with 6 wavenumbers with randomized phase and the same initial energy in the balanced and wave modes, \tit{i.e.}, $E_{t0}/E_{w0} = 1$. $\Ro = \Bu = 0.01$.}
\label{simTrandomweakwaves}
\end{figure}

\begin{figure}
\centering
\includegraphics[keepaspectratio=true,width=0.5\textwidth]{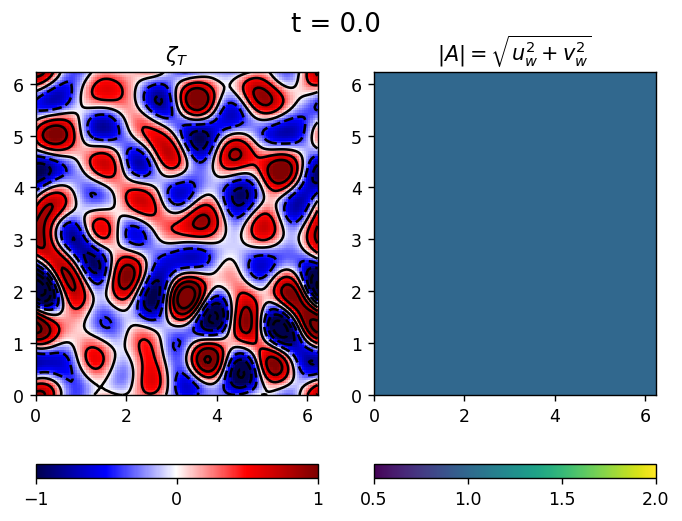}\includegraphics[keepaspectratio=true,width=0.5\textwidth]{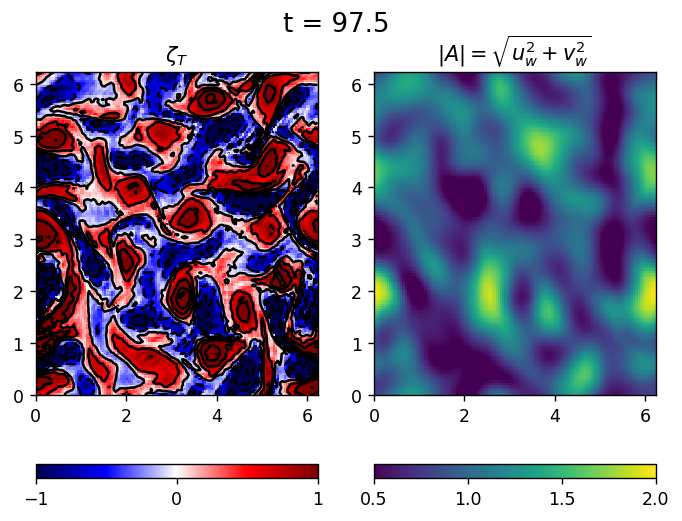}\\
\includegraphics[keepaspectratio=true,width=0.7\textwidth]{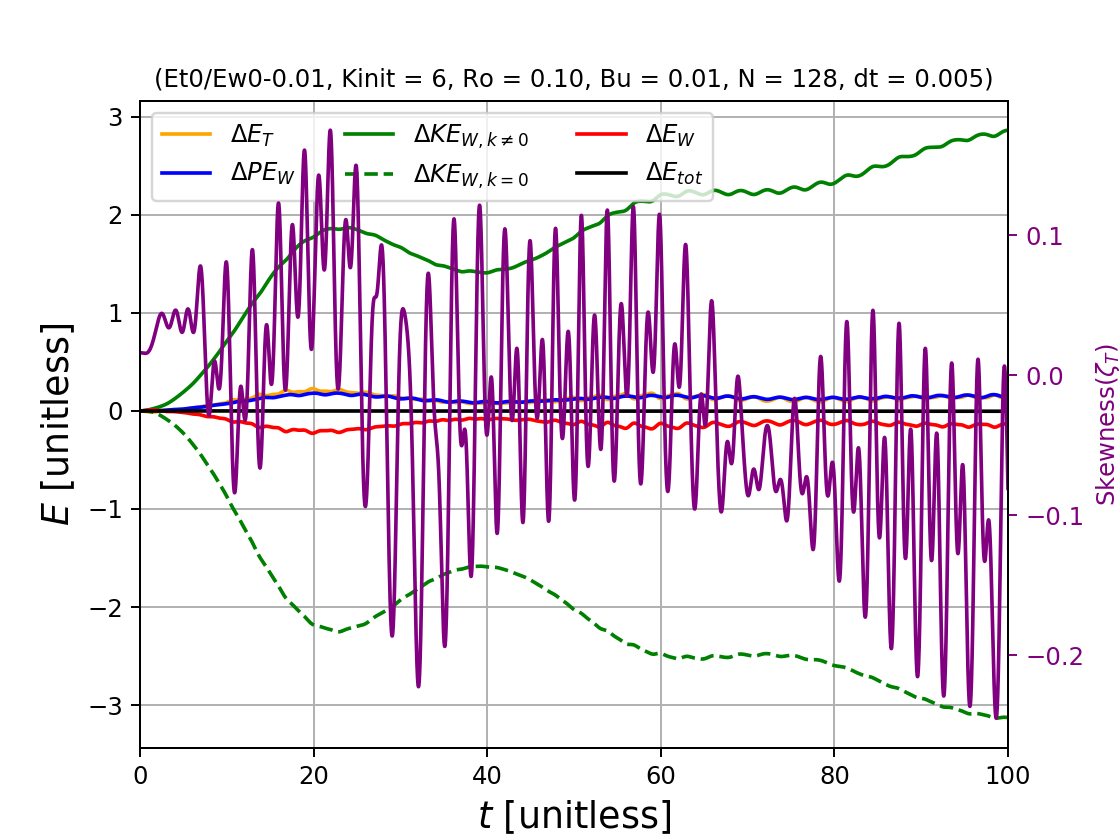}
\caption{Same as Figure \ref{simTWanticyclone}, but initialized with 6 wavenumbers with randomized phase, and a hundred times more initial energy in the wave modes, \tit{i.e.}, $E_{t0}/E_{w0} = 0.01$. $\Ro = 0.1$ and $\Bu = \Ro^2 = 0.01$).}
\label{simTrandomstrongwaves}
\end{figure}

When the initial conditions have both anticyclones and cyclones randomly distributed across a few low wavenumbers, the behavior is qualitatively similar to when only one sign of vorticity is initially present, in the sense that anticyclones trap wave energy while cyclones repel it (upper-right panels of Figures \ref{simTrandomweakwaves} and \ref{simTrandomstrongwaves}). The energy changes are also similar, with the waves losing total energy while the balanced flow gains total energy. The wave energy is initially contained entirely in the purely inertial, spatially uniform mode ($k = 0$), but it decreases rapidly mirroring the increase in wave energy in the higher modes ($k \neq 0$). Importantly, the energy changes appear to be relatively insensitive to the relationship between $\Ro$ and $\Bu$ and the initial balanced/wave energy ratio, $E_{t0}/E_{w0}$ (compare Figure \ref{simTrandomweakwaves}, where $\Bu = \Ro = 0.01$ and $E_{t0}/E_{w0} = 1$, with Figure \ref{simTrandomstrongwaves}, where $\Bu = \Ro^2 = 0.01$ and $E_{t0}/E_{w0} = 0.01$). The vortices in the simulation where $E_{t0}/E_{w0} = 0.01$ are more deformed, with a less smooth vorticity distribution (compare upper-right panels of Figures \ref{simTrandomweakwaves} and \ref{simTrandomstrongwaves}), indicating that the balanced flow can be appreciably impacted by the near-inertial waves in this strong wave regime.

\clearpage
\section{Asymptotic model for NIWs-balanced flow interaction}

In this section we derive a new asymptotic model that represents both clockwise and counterclockwise wave modes. We begin with the truncated equations derived by \cite{thomas_etal2017}, linearized about a steady balanced barotropic flow, written in complex representation:

\beq
\U_t + i\U + 2\Bu p_{\scc} + \eps F^u = 0,\\
p_t + \U_s + \overline{\U}_{\scc} + \eps F^p = 0,
\eeq

\nin where

\beq
F_u \equiv \U_T + U\U_s + \overline{U}\U_{\scc} + \frac{i}{2}\big(\U\zeta + \overline{\U}\sigma\big),\\
F_p \equiv p_T + Up_s + \overline{U}p_{\scc},\\
\zeta \equiv \triangle\Psi, \sigma \equiv \triangle^\perp\Psi
\eeq

\nin Following \cite{thomas_etal2017}'s Appendix B, the governing equations can be rewritten only in terms of velocity in complex representation as

\beq
\partial_t\big(\partial^2_{tt} + 1 - 4\Bu\partial^2_{s\scc}\big)\U + \eps R^u = 0\\
\partial_t\big(\partial^2_{tt} + 1 - 4\Bu\partial^2_{s\scc}\big)p + \eps R^p = 0,
\eeq

\nin where

\beq
R^u \equiv iF^u_t - F^u_{tt} + i\Bu\big(F^u_{s\scc} - F^u_{ss}\big) + i\Bu\big(\overline{F^u_{s\scc}} - \overline{F^u_{\scc\scc}}\big) + 2\Bu\big(F^p_{\scc t} - iF^p_{\scc}\big)\\
R^p \equiv F^u_{st} + \overline{F^u_{\scc t}} + i\big(F^u_{\scc} - F^u_s\big) - F^p_{tt} - F^p.
\eeq

\nin We write the solutions as

\beq
\U = A^-e^{-i\omega t} + A^+e^{+i\omega t},\\
p = - \frac{i}{\omega}\big(A^-_s + \overline{A^+}_{\scc}\big)e^{-i\omega t} + \frac{i}{\omega}\big(A^+_s + \overline{A^-}_{\scc}\big)e^{+i\omega t}
\eeq


\nin Expanding $\U$ and $p$ in powers of $\eps$:

\beq
\U = \U^{(0)} + \eps\U^{(1)} + \eps^2\U^{(2)} + \cdots
\eeq

\nin and substituting in the momentum equation,

\beq
i\omega(\mp\omega^2 \pm 1 \mp 4\Bu\partial^2_{s\scc})A_0^\mp &  & = 0\\
\underbrace{i\omega(\mp\omega^2 \pm 1 \mp 4\Bu\partial^2_{s\scc})}_{\equiv \mathcal{M}^\mp}A_1^\mp & + R_0^\mp & = 0
\eeq

\nin To obtain a single pair of equations for $A^\mp \equiv A_0^\mp + \eps A_1^\mp$, we follow the reconstitution technique as used by \textit{e.g.}, \cite{roberts1985,wagner_young2016,thomas2017}. The first step is to add a small correction to the RHS of the $O(1)$ equations:

\beq
\mathcal{M}^\mp A_0^\mp &  & = \eps\Phi^\mp\label{rec1}\\
\mathcal{M}^\mp A_1^\mp & + R_0^\mp & = 0\label{rec2}
\eeq

\nin Take (\ref{rec1}) + $\Ro\times($\ref{rec2}):

\beq
\Ro\Phi^\mp = -\Ro\big[\mathcal{M}^\mp A_1^\mp - R_0^\mp\big]
\eeq

\nin Substitute back in \ref{rec1}:

\beq
\mathcal{M}^\mp A^\mp + \Ro R_0^\mp = 0
\eeq

\nin After some manipulations, the coupled equations for $A^-$ and $A^+$ become:

\beq\label{coupledamp}
\big[\mathcal{L}^\mp - 2\Bu(1 + i)\partial^2_{s\scc}\big]A^\mp_T = 2\Bu(1 + i)(\overline{A^\pm}_{T\scc\scc} + \beta^\mp) - \mathcal{L}^\mp\alpha^\mp \cdots\\
\cdots \mp \frac{i\omega}{\eps}\big(-\omega^2 + 1 - 4\Bu\partial^2_{s\scc}\big)A^\mp + D A^\mp,
\eeq

\nin where the hyperviscosity operator $D \equiv \nu\triangle^{2r}$ has been added and

\beq
\mathcal{L}^\mp \equiv \big\{\omega(\omega \pm 1) + i\Bu\big[\partial^2_{s\scc} - \partial^2_{ss} + \partial^2_{s\scc}(\overline{\bullet}) - \partial^2_{\scc\scc}(\overline{\bullet})\big]\big\},\\
\alpha^\mp \equiv UA^\mp_s + \overline{U}A^\mp_{\scc} + \frac{i}{2}\big(A^\mp\zeta + \overline{A^\pm}\sigma\big),\\
\beta^\mp \equiv \big[U(A^\mp_{ss} + \overline{A^\pm}_{s\scc}) + \overline{U}(A^\mp_{s\scc} + \overline{A^\pm}_{\scc\scc})\big]_{\scc}.
\eeq

\nin in Cartesian coordinates, Equation \ref{coupledamp} reads

\beq
\bigg[\mathcal{L}^\mp - \frac{1}{2}\Bu(1 + i)\triangle\bigg]A^\mp_T = 2\Bu(1 + i)\bigg(\frac{1}{4}\triangle^\perp\overline{A^\pm}_T + \beta^\mp\bigg) - \mathcal{L}^\mp\alpha^\mp \cdots\label{eqampcart}\\
\cdots \mp \frac{i\omega}{\eps}\big(-\omega^2 + 1 - \Bu\triangle\big)A^\mp + D A^\mp,\nonumber
\eeq

\nin and

\beq
\mathcal{L}^\mp \equiv \omega(\omega \pm 1) + \frac{i}{2}\Bu\big(\partial^2_{yy} + i\partial^2_{xy}\big) + \frac{i}{2}\Bu\big(\partial^2_{yy} - i\partial^2_{xy}\big)(\overline{\bullet}).
\eeq

\nin If $\omega = 1$ and $\Bu$, $A^+ \to 0$, the YBJ \cite{young_benjelloul1997} amplitude equation is recovered for $A^-$:

\beq
\boxed{
A^-_T + \bm{U\cdot\nabla}A^- + \frac{i}{2}\bigg(A^-\zeta - \frac{\Bu}{\eps}\triangle A^-\bigg) = 0.
}
\eeq

\subsection{Numerical implementation}

Our next goal is to solve \ref{eqampcart} numerically. Equation \ref{eqampcart} can be rewritten as

\beq\label{eqampcart2}
\bigg[(\omega \pm 1) - \frac{\Bu}{2}\tilde{\triangle}\bigg]A^\mp_T - \frac{\Bu}{2}(1 + i)\triangle^\perp\overline{A^\pm_T} \pm i\omega(-\omega^2 + 1 - \Bu\triangle)A^\mp - DA^\mp = \cdots\nonumber\\
\cdots - \mathcal{L}^\mp\alpha^\mp + 2\Bu(1 + i)\beta^\mp,
\eeq

\nin and semi-discretized with a forward-in-time scheme as

\beq\label{eqampcart3}
\bigg[\omega(\omega \pm 1) - \frac{\Bu}{2}\tilde{\triangle}\bigg]\frac{A^{\mp(n+1)} - A^{\mp(n)}}{\eps\delta t} - \frac{\Bu}{2}(1 + i)\triangle^\perp\frac{\overline{A^{\pm(n+1)}} - \overline{A^{\pm(n)}}}{\eps\delta t} \cdots\nonumber\\
\cdots \pm i\omega(-\omega^2 + 1 -\Bu\triangle)A^{\mp(n+1)} - DA^{\mp(n+1)} = -\mathcal{L}^\mp\alpha^{\mp(n)} + 2\Bu(1 + i)\beta^\mp
\eeq

\nin Multiplying through by $\eps\delta t$ and rearranging yields:

\beq
\bigg[\omega(\omega \pm 1) - \frac{\Bu}{2}\tilde{\triangle} \pm i\omega\eps\delta t(-\omega^2 + 1 - \Bu\triangle) - \eps\delta tD\bigg]A^{\mp(n+1)} - \frac{\Bu}{2}(1 + i)\triangle^\perp \overline{A^{\pm(n+1)}} = \cdots \nonumber\\
\cdots \bigg[\underbrace{\omega(\omega \pm 1) - \frac{\Bu}{2}\tilde{\triangle}}_{\equiv\tilde{\mathcal{L^\mp}}}\bigg]A^{\mp(n)} - \frac{\Bu}{2}(1 + i)\triangle^\perp \overline{A^{\pm(n)}} + \eps\delta t\big(2\Bu(1 + i)\beta^{\mp(n)} - \mathcal{L}^\mp\alpha^{\mp(n)}\big),
\eeq

\nin where

\beq
\tilde{\triangle} \equiv \partial^2_{xx} + \partial^2_{yy} + \partial^2_{xy} + i\partial^2_{xx}
\eeq

\nin and the $(n)$ and $(n+1)$ superscripts indicate the time step at which the term is evaluated. Writing $A^\mp$ in terms of their real and imaginary parts, \textit{i.e.}, $A^\mp = A^\mp_R + iA^\mp_I$, we have

\beq\label{eqampcart4}
\bigg[\mathcal{\tilde{L^\mp}} \pm i\omega\eps\delta t(-\omega^2 + 1 - \Bu\triangle) - \eps\delta tD\bigg]\big(A^\mp_R + iA^\mp_I\big)^{(n+1)} - \frac{\Bu}{2}(1 + i)\triangle^\perp\big(A^\pm_R - iA^\pm_I\big)^{(n+1)} = \cdots\nonumber\\
\cdots\mathcal{\tilde{L^\mp}}\big(A^\mp_R + iA^\mp_I)^{(n)} - \frac{\Bu}{2}(1 + i)\triangle^\perp\big(A^\pm_R - iA^\pm_I)^{(n)} + \eps\delta t\big(2\Bu(1 + i)\beta^{\mp(n)} - \mathcal{L}^\mp\alpha^{\mp(n)}\big),
\eeq

\nin Equation \ref{eqampcart4} and its complex conjugate can be written respectively as

\beq
\mathcal{H}^\mp_1A_R^{\mp(n+1)} + i\mathcal{H}^\mp_1A^{\mp(n+1)}_I + \mathcal{H}_2A_R^{\pm(n+1)} - i\mathcal{H}_2A^{\pm(n+1)}_I = R^{\mp(n)}
\eeq

\nin and

\beq
\overline{\mathcal{H}^\mp_1}A_R^{\mp(n+1)} - i\overline{\mathcal{H}^\mp_1}A^{\mp(n+1)}_I + \overline{\mathcal{H}_2}A_R^{\pm(n+1)} + i\overline{\mathcal{H}_2}A^{\pm(n+1)}_I = \overline{R^{\mp(n)}}.
\eeq

\nin In matrix form,

\begin{equation}
\begin{bmatrix}
    \mathcal{H}_1^- & i\mathcal{H}_1^- & \mathcal{H}_2 & -i\mathcal{H}_2 \\
    \mathcal{H}_2 & -i\mathcal{H}_2 & \mathcal{H}_1^+ & i\mathcal{H}_1^+ \\
    \overline{\mathcal{H}_1^-} & -i\overline{\mathcal{H}_1^-} & \overline{\mathcal{H}_2} & i\overline{\mathcal{H}_2} \\
    \overline{\mathcal{H}_2} & i\overline{\mathcal{H}_2} & \overline{\mathcal{H}_1^+} & -i\overline{\mathcal{H}_1^+} \\
\end{bmatrix}
\begin{bmatrix}
    A^{-(n+1)}_R \\
    A^{-(n+1)}_I \\
    A^{+(n+1)}_R \\
    A^{+(n+1)}_I
\end{bmatrix}
=
\begin{bmatrix}
    R^{-(n)} \\
    R^{+(n)} \\
    \overline{R^{-(n)}} \\
    \overline{R^{+(n)}}
\end{bmatrix}
\end{equation}

\nin The next step is to implement and time-step this system numerically.

\section{Conclusions}

The main results of this project are as follows:

\begin{itemize}
\item{A new two-component 2D model seems to reproduce the energetics of more complex 3D models, \textit{i.e.}, \textbf{energy transfer from NIWs to the balanced flow at low $\Ro$};}
\item{This might be the simplest non-asymptotic two-component model that also captures \textbf{balanced energy dissipation at high $\Ro$};}
\item{Idealized models such as the ones developed in this study can be used as \textbf{testbeds for parameterizations} for global 3D models in regions of high NIW energy (instead of artificially enhanced viscosity) and}
\item{A \textbf{new asymptotic model} was derived for NIW-balanced flow interactions.}
\end{itemize}

\section{Next steps}

The next steps in this project are to

\begin{itemize}
\item{Verify if the \tbf{direction of the energy exchange} (NIWs$\rightarrow$balanced) changes as $\Ro$ = O(1) is approached (computationally demanding) and}
\item{\tbf{Test the new asymptotic model} against the parent T-W model.}
\end{itemize}

\section*{Acknowledgements}

It is a pleasure to acknowledge the work of the directors, Claudia Cenedese, Bruce Sutherland and Karl Helfrich, in running a wonderful and inspiring GFD summer. Thanks to the other fellows and the GFD staff, visitors and seminar speakers for stimulating and insightful discussions. Many thanks to Jim Thomas for proposing and patiently supervising this project.

\bibliography{abbreviations,references}
\end{document}